\documentclass[iop]{emulateapj}

\usepackage[english]{babel}
\usepackage[utf8x]{inputenc}
\usepackage[T1]{fontenc}

\usepackage{amsmath,bm}
\usepackage{amssymb}
\usepackage{graphicx}
\usepackage[colorlinks=true, allcolors=blue]{hyperref}
\usepackage{booktabs}
\usepackage{ulem}
\usepackage{color}
\usepackage{amsmath}
\usepackage[symbol]{footmisc}
\usepackage{natbib}

\shorttitle{Efficiency of Planetesimal Accretion - II: Effect of Saturn}
\shortauthors{Haghighipour, Podolak, Podolak}

\begin{document}

\title{Detailed Calculations of the Efficiency of Planetesimal Accretion in the Core-Accretion Model -II:
The effect of Saturn}

\author{Nader Haghighipour\altaffilmark{1,2,3}, Morris Podolak\altaffilmark{4}, and Esther Podolak\altaffilmark{5}}

\altaffiltext{1}{Planetary Science Institute, Tucson, AZ 85719, USA} 
\altaffiltext{2}{Institute for Astronomy, University of Hawaii-Manoa, Honolulu, HI 96822, USA}
\altaffiltext{3}{Institute for Advanced Planetary Astrophysics, Honolulu, HI, USA}
\altaffiltext{4}{Department of Geosciences, Tel Aviv University, Tel Aviv, Israel 69978}
\altaffiltext{5}{Liacom, Holon, Israel}

\begin{abstract}
As part of our ongoing initiative on accurately calculating the accretion rate of planetesimals 
in the core-accretion model, we demonstrated in a recent article that when the calculations include 
the gravitational force of the Sun (the original core-accretion model did not include solar gravity), 
results change considerably [ApJ, 899:45]. In this paper, we have advanced our previous study by including 
the effect of Saturn. To maintain focus on the effect of this planet, and in order to be consistent with previous 
studies, we did not include the effect of the nebular gas. Results demonstrated that as expected, 
Saturn's perturbation decreases the rate of accretion by scattering many planetesimals out of Jupiter's 
accretion zone. It also increases the velocities with which planetesimals encounter the envelope, which 
in agreement with our previous findings, enhances their break-up due to the ram-pressure. Results also 
show that, because the effect of Saturn in scattering of planetesimals increases with its mass, this planet 
might not have played a significant role in the accretion of planetesimals by proto-Jupiter during the 
early stage of its growth. Finally, the late accretion of planetesimals, as obtained in our previous 
study, appears in our new results as well, implying that combined with the rapid in-fall of the gas, 
it can result in the mixing of material in the outer regions of the envelope which may explain the 
enhancement of the envelope's high-Z material.
\end{abstract}

\section{Introduction}
It is generally accepted that the collisional growth of km-sized planetesimals serves as a prelude to the
formation of planetary systems. 
Depending on their radial distances from the central star, which determines their compositions,
material strength, and impact velocities, the collision and accretion of these bodies can result in the formation of 
Moon- to Mars-sized planetary embryos in the inner part of a protoplanetary disk, and the cores of giant planets 
in regions beyond the disk's snowline. The core-accretion model of giant planet formation suggests that
the latter is naturally followed by the accretion of gas from the nebula and results in the 
formation of gas-giant planets. 

During the formation of a gas-giant, the contribution of planetesimals continues beyond the formation 
of the core. In the gas-accretion phase, and also during the time that the protoplanet's envelope collapses, 
many planetesimals enter the envelope and contribute to its metallicity and the luminosity of the planet 
by depositing their materials as they vaporize due to heating by the gas-drag. 

While the physics of processes governing planetesimal-envelope interactions are independent 
of the characteristics of the system, their outcomes directly depend on the physical and 
dynamical properties of planetesimals and the state of the gas. For instance, as the envelope evolves and 
its density and temperature structures change, its response to encountered planetesimals varies which
causes the rate of the accretion of these objects to vary as well. 
These variations in the accretion rate of planetesimals strongly affect
the formation of the planet as the outcome of the accretion determines the composition of the envelope 
and the onset of its collapse \citep{Pollack96,Iaros07,Movsh10,Dangelo14,Venturini16,Lozovsky17}.
It is therefore, imperative that any model of giant planet formation take these interactions into account
and calculate their contributions accurately and self-consistently. 

Because the strength of the interaction of planetesimals with the envelope, which determines the efficacy
of their vaporization and mass-deposition, depends directly on their encounter velocities, to calculate
the realistic amount of the mass that is deposited during the evolution of the envelope, 
the rate of the encounters of planetesimals and their encounter velocities must be obtained from the natural 
dynamical evolution of the system. A comprehensive model will also need to continue the integration of
those planetesimals that enter the envelope, and calculate their mass-deposition in concert with the 
envelope's density and temperature variations. 

While the above requirements have been known since the inception of the core-accretion model, realistic 
simulations have been possible only during the past few years. For almost three decades, limitations in
computer technology did not allow for large scale simulations, forcing scientists to tune their approaches 
to the available computational resources by making simplifying assumptions. For instance, in the original work
of \cite{Pollack96}, planetesimal contributions were calculated using a semi-analytical approach. 
To avoid computational complexities, these authors assumed that all planetesimals entered 
the envelope with the same velocity and determined the rate of their encounters using a gravitational enhancement
factor that was obtained from fitting to the results of the numerical simulations of \citet{Greenzweig90,Greenzweig92}. 
Once inside the envelope, the orbit of a planetesimal and the amount of its mass-deposition were calculated using 
the methodology and the code developed by \citet{Podolak88}. The orbit was, however, integrated including
only the effect of the proto-Jupiter (i.e., a two-body system) and external sources such as the gravitational
forces of the Sun and additional (growing) giant planet(s) were not included. \citet{Pollack96} considered a 
planetesimal to be fully accreted if its total energy became lower than a minimum value, or if it broke 
up due to ram-pressure. The mass-accretion rate was then determined by carrying out the above calculations for 
a series of impact parameters and calculating the protoplanets effective capture cross section and planetesimals 
flux.

Subsequent advances in computational techniques combined with the desire to achieve a shorter formation time
for Jupiter prompted researchers to revisit some of the above processes. For instance, \citet{Inaba03} examined 
the full-accretion criteria by developing an analytical approximation to the time-evolution of the envelope 
as well as an analytical fit to the results of the N-body simulations of \citet{Ida93}, \citet{Nakazawa89}, 
and \citet{Inaba01}. These authors ignored the effect of the Sun on the orbits of planetesimals 
inside the envelope and calculated the ablation of planetesimals using an ablation factor that
did not vary with the envelope's temperature. \citet{Inaba03} argued that their simulations point to a larger rate of 
planetesimal encounters and, therefore, a shorter formation time for Jupiter. They also argued that
based on their results, ram-pressure would not reach high values to break up planetesimals. 

While the above calculations demonstrated the proof of the concept, their results were not fully realistic.
In the work of \citet{Pollack96}, the encounter flux of planetesimals and their velocities had not been obtained 
from the natural dynamical evolution of the system, and the analytical approach of \citet{Inaba03}
ignored the effect of the variation of the envelope's temperature 
on the ablation of planetesimals. Furthermore, neither of the above calculations included the
gravitational perturbation of the Sun and other giant planets. 

In a recent article \citep[][hereafter, Paper-I]{Podolak20}, we revisited the planetesimal-envelope interaction
and developed a comprehensive and self-consistent approach that would address 
the above shortcomings. Using a special purpose integrator (ESSTI, see section 2.4 for more details), 
we integrated the orbits of a large number of planetesimals starting from their original locations in the 
protoplanetary disk and calculated the ablation of those that entered the envelope using their encounter 
velocities obtained from their orbital evolution. Our calculations took into account the time-variations 
of the density and temperature of the envelope, and as the first step to address some of the shortcomings 
of the previous studies, they included the gravitational effect of the Sun as well.

\begin{figure}[ht]
\vskip -15pt
\hskip -25pt
\includegraphics[scale=0.41]{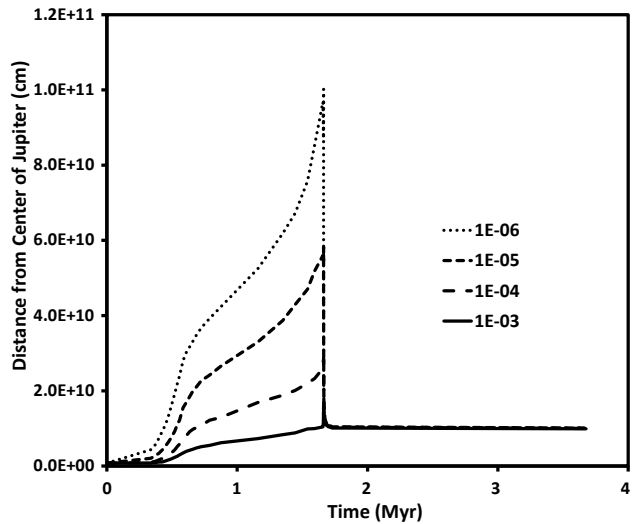}
\vskip -30pt
\caption{The graphs of the time-varations of the density of the envelope at different distances from the center of Jupiter
based on the models described by \citet{Lozovsky17}. Shown are gas densities of ${10^{-6}}\, {\rm g \, {cm^{-3}}}$ 
(dotted line), ${10^{-5}}\, {\rm g \, {cm^{-3}}}$ (small-dashed line), ${10^{-4}}\, {\rm g \, {cm^{-3}}}$ (large-dashed line), 
and ${10^{-3}}\, {\rm g \, {cm^{-3}}}$ (solid line).}
\label{fig1}
\end{figure}

In this study, we advance the calculations of Paper-I by including the gravitational perturbation of a planet
in the orbit of Saturn. As demonstrated by \citet{Haghighipour12} and \citet{Haghighipour16}, Saturn played
a significant role in the formation of terrestrial planets and the structure of asteroid belt. It also 
contributed significantly to the origin of the parent bodies of the iron meteorites \citep{Haghighipour12}.
Given the extent of the perturbation of this planet, it is certain that Saturn has also played a significant role 
in the orbital evolution of planetesimals in the region beyond 5 AU, especially those in the influence zone 
of Jupiter. It is expected that the gravitational perturbation of Saturn directly influenced the encounter
rate of planetesimals as well as the velocity with which they entered Jupiter's envelope. In this paper, 
we calculate these effects and study their consequences on planetesimal-envelope interaction and the metalicity 
of the envelope.

The rest of this paper is organized as follows. In Section 2, we explain our numerical approach and describe the details
of our system and its initial set up. In this section, we also discuss the physics of planetesimal-envelope
interaction and the way that they have been included in our calculations. We present the results in Section 3
and discuss the effect of Saturn in Section 4. We conclude our study in Section 5 by summarizing our findings
and discussing their implications for the formation of gas-giant planets in our solar system and beyond.

\begin{deluxetable*}{llll}
\tablecaption{Physical parameters for different compositions.}
\tablehead{
\colhead{\hskip -15pt Planetesimal Parameters} & 
\colhead{\hskip -21pt Pure Ice} &  
\colhead{\hskip -18pt Pure Rock}  & 
\colhead{\hskip -31pt Mixed} 
}
\startdata
Bulk density $({\rm g cm^{-3}})$   & 1.0                   &  3.4                   &  2.0                   \\
Tensile Strength $({\rm dyn\, cm^{-2}})$ & $10^6$       &   $10^8$       &    $10^6$        \\
$T_{\rm crit}$ $({\rm K})$       & 648                   & 4000                    & 648                      \\
$m_{\rm p}$ $({\rm g})$       & $2.99\times 10^{-23}$  &  $8.31\times 10^{-23}$  &  $2.99\times 10^{-23}$  \\
$E_0$ $({\rm erg})$           & $2.8\times 10^{10}$    & $8.08\times 10^{10}$    &  $2.8\times 10^{10}$     \\
$P_0$ $({\rm dyn\, cm^{-2}})$           & $3.891\times 10^{11}$  & $1.50\times 10^{13}$    &  $3.891\times 10^{11}$   \\
$A$ $({\rm K})$             & $-2.1042\times 10^3$  & $-2.4605\times 10^4$    & $-2.1042\times 10^3$    \\
\enddata
\end{deluxetable*}

\section{The Initial Set up and Numerical Approach}

\subsection{The System}
We consider a four-body system consisting of the Sun, Jupiter, Saturn and a planetesimal. The mass of the 
planetesimal is negligible and its orbit is between Jupiter and Saturn. Because our goal is to determine 
the amount of the mass that a planetesimal loses during its passage through the Jupiter's envelope, we include 
the gravitational force of Saturn as a constant perturbation, and treat Jupiter as an extended mass. 
For the purpose of this study, we do not consider a diffused core in Jupiter. Instead, we assume that the
core is condensed and has a well-defined surface, separating heavy elements and H/He.

During the evolution of the system, the variations of the radius, mass, and internal density 
distribution of Jupiter are included using the model A of \citet{Lozovsky17}. This model follows the growth 
of a planet in the core-accretion scenario for an assumed background surface density of solids of 
$\sigma=6\, {\rm {g \, cm^{-2}}}$ and a planetesimal radius of 100\,km. Figure 1 shows the variations of
the envelope's gas density in one of such models. The dotted line represents
the gas density of ${10^{-6}}\,{\rm g \, cm^{-3}}$, small and large dashed lines correspond to the gas densities of 
${10^{-5}}\,{\rm g \, cm^{-3}}$ and ${10^{-4}}\,{\rm g \, cm^{-3}}$, respectively, and the solid line shows a gas density of
${10^{-3}}\,{\rm g \, cm^{-3}}$. As shown by the figure, when Jupiter grows, its envelope grows as well, 
both in radius and mass. This causes the density of the gas to vary with time at different elevations inside the envelope. 
In this model, at approximately $1.7 \times {10^6}$ years, the envelope becomes unstable and the runaway 
gas-accretion ensues. 

To avoid complications due to the time evolution of the nebula, and for the mere purpose of maintaining 
focus on the perturbing effects of Saturn, we do not include the interaction of planetesimals with the 
nebular gas in our calculations. We also do not consider the appearance of a gap in the gaseous disk around 
the orbit of proto-Jupiter \citep{Shibata19} and the migration of the planet due to its interaction with the 
nebula \citep{Shibata20,Shibata22}. While these effects are indeed important, they strongly depend
on the model used for the disk gas distribution. They also depend on the subsequent evolution of the gas disk 
over the 3 Myr period we are considering here. Because, at present, there is no standard model for the structure 
and evolution of the disk, we prefer to focus on the dynamical effect of a growing Saturn, and put aside, 
for the present, the disk-model-dependent effects.

\subsection{The Composition and Structure of Planetesimals}
Planetesimals are considered to be perfectly spherical and have three sizes of 1, 10, and 100 km.
Because during the interaction of a planetesimal with Jupiter's envelope, the efficiency by which it 
loses mass, and the amount of its mass-loss, in addition to its size, shape and velocity, vary with
its internal composition as well, we consider three compositions for each size of the planetesimals:

\vskip 5pt
\noindent
{\bf Pure ice}. With a bulk density of $1.0 \, {\rm g \, cm^{-3}}$, these planetesimals represent the low-end
of the density spectrum. Pure-ice planetesimals are volatile and evaporate easily as they are heated up by gas-drag
 while passing through the envelope. 

\vskip 5pt
\noindent
{\bf Pure rock}. These planetesimals represent the hardest material with the bulk density of $3.4 \, {\rm g \, cm^{-3}}$. 
They are not easily evaporated and their dynamics is less strongly affect by gas-drag. Rocky
planetesimals also have high tensile strength which makes them stable against break-up due to ram-pressure.

\vskip 5pt
\noindent
{\bf Mix of ice plus rock}. In reality, a planetesimal in the region between Jupiter and Saturn is made of a mix of
ice and rock. 
When such a planetesimal is heated, its ice component vaporizes and as it escapes the planetesimal,
it carries some of its rocky material with it. This rocky debris enhances the metalicity of the envelope 
and increases its dust/gas ratio. As the ice-vapor escapes the planetesimal, it also increases the
planetesimal's porosity weakening its internal strength which in turn makes the planetesimal more susceptible to 
breakage due to ram-pressure. It is, therefore, imperative to study the response of such mixed planetesimals to
their interactions with envelope as these bodies can have large contributions to the envelope's metalicity and the 
possible growth of the planet's core. In choosing the fraction of the ice and rock in our mix-composition planetesimals, 
we follow the trend observed in the bulk density of some of the large bodies in the outer 
solar system and consider planetesimals with 30\% ice and 70\% rock corresponding to a bulk density of 
$2.0 \, {\rm g \, cm^{-3}}$. This bulk density is similar to that of objects such as Pluto, Triton, and Titan,

\subsection{Planetesimal-Envelope Interaction} 
When inside the envelope, in addition to the gravitational forces of the Sun, Saturn, Jupiter's core and 
the portion of the envelope that is interior to its orbit, the motion of a planetesimal is also affected 
by gas-drag. At this stage, the combined heating due to the gas-drag and the radiation received 
from the hot ambient gas, increases the temperature of the planetesimal causing it to lose mass.
The rate of mass-loss depends on the material composition of the planetesimal and its response to heating.
The latter is quantified by the critical temperature of the planetesimal material $(T_{\rm crit})$ above which there 
is no phase change. 

If, while the planetesimal passes through the envelope, its surface temperature $(T_{\rm p})$ does not 
exceed $(T_{\rm crit})$, the surface material can exist in two phases of solid and gas (vapor). At this
state, the mass-loss will be primarily due to evaporation,

\begin{equation}
{{d{M_{\rm p}}}\over{dt}} = - 4 \pi {R_{\rm p}^2} {P_{\rm vap}}
\Bigl({{\pi {m_{\rm p}}}\over{8k{T_{\rm p}}}}{\Bigr)^{1/2}}\,.
\end{equation}

\noindent
In this equation, $M_{\rm p}$ and $R_{\rm p}$ are the mass and radius of the planetesimal, $m_{\rm p}$ 
is the mass of the planetesimal molecule, and $k$ is the Boltzmann's constant. The quantity $P_{\rm vap}$
given by

\begin{equation}
{P_{\rm vap}} = {P_0}\, {e^{-A/{T_{\rm p}}}}\,,
\end{equation}

\noindent
represents the upper limit of the pressure of the vapor surrounding the planetesimal. If the vapor 
pressure exceeds $P_{\rm vap}$, any excess vapor re-condenses onto the planetesimal surface until 
its pressure stays equal or drops below the above value. In equation (2), $P_0$ and $A$ are constant 
quantities whose values depend on the composition of the planetesimal. Tables 1 shows the values 
of these quantities and other properties of the planetesimals used in our study.

Once the planetesimal's surface temperature reaches $T_{\rm {crit}}$, the vapor can have any pressure.
We assume that at this state, any surface material is automatically turned into vapor and is blown away by 
the ambient gas that streams past of it. The mass-loss will then be due to the combined effects 
of evaporation and radiation heating/cooling, and is given by

\begin{equation}
{{d{M_{\rm p}}}\over{dt}} = {{4 \pi {R_{\rm p}^2}}\over{E_0}}
\Bigl[{1\over {16}}{C_{\rm D}}{\rho_{\rm g}}{v_{\rm rel}^3}+\sigma({T_{\rm g}^4}-{T_{\rm crit}^4})\Bigr]\,.
\end{equation}

\noindent
In this equation, $\rho_{\rm g}$ and $T_{\rm g}$ are the density and temperature of the envelope gas 
at the position of the planetesimal, $v_{\rm rel}$ represents the magnitude of the velocity of the 
planetesimal relative to the gas $({v_{\rm rel}}>0)$, $\sigma$ is the Stefan-Boltzmann constant, $C_{\rm D}$ is the drag 
coefficient and $E_0$ is a constant quantity whose value depends on the composition of the planetesimal
(see Table 1). We refer the reader to the appendix of Paper-I and \citet{Podolak88} for more details
on the derivations of these equations.

\subsection{Numerical Integrations}
To accurately quantify the interaction of a planetesimal with the envelope, specifically, its mass-loss
and size reduction, and to ensure that the velocity with which the planetesimal enters the 
envelope would be the natural outcome of its dynamical evolution, we integrated the orbit of the planetesimal 
starting from its initial position between Jupiter and Saturn.

To avoid complications due to the stiffness of the differential equations, we broke the integrations into two parts:
inside and outside the envelope. When outside the envelope, we integrated the four-body system of 
Sun-Jupiter-planetesimal-Saturn using the $N$-body integrator
MERCURY \citep{Chambers99}. At this stage, we included Saturn as a point mass and Jupiter as
an extended object whose radius and mass varied with time according to the model A of \citet{Lozovsky17}. When
the integrator indicated a collision between the planetesimal and the extended-mass Jupiter, $N$-body integrations
were stopped and integrations were continued using our special purpose integrator, ESSTI (Explicit Solar System 
Trajectory Integrator). 

ESSTI has been developed to integrate the orbit of a planetesimal inside the proto-Jupiter's envelope and outside,
at any location in a protoplanetary disk. It has been designed to automatically include in the equations of motion, 
the physical processes associated with the location of the planetesimal. For instance, 
when a planetesimal enters the envelope, ESSTI adds to the equations of motion the drag force of the gas. 
It keeps the forces of the Sun and Saturn intact, but replaces the gravitational force of Jupiter with 
the gravitational force of the portion of the mass that is located between the orbit of the planetesimal 
and the center of mass of Jupiter. It accounts for the planetesimal's mass-loss by including the effects of 
the aerodynamical heating due to gas-drag, radiative heating due to the ambient gas, 
radiative cooling, and evaporation cooling. All these processes have been hard-wired in ESSTI with their
corresponding analytical formulae (Table 2). We refer the reader to Paper-I for more details on this integrator.

In addition to accounting for mass-loss due to ablation, ESSTI also includes the effect of ram-pressure. 
The latter is due to the pressure difference across the planetesimal caused by the balk motion 
of the gas. For a planetesimal moving with a relative velocity $v_{\rm rel}$ in a gas with a density 
$\rho_{\rm g}$, the ram-pressure on its front hemisphere is given by \citep{Pollack79}

\begin{equation}
{P_{\rm ram}} = {1\over 2}\,{\rho_{\rm g}}\,{v_{\rm rel}^2}\,.
\end{equation}

\noindent
The planetesimal will fragment when the ram-pressure exceeds the object's compressive strength. 
If the self-gravity of the fragmented body cannot hold the fragments together, the body will
break apart. The critical radius above which the self-gravity of the planetesimal will be strong enough 
to hold its fragments together is \citep{Pollack79}

\begin{equation}
{R_{\rm p \, (self-grav)}} = {1\over 2}\, {v_{\rm rel}} \, {\Bigl({{\rho_{\rm g}}\over {\rho_{\rm p}}}\Big )}\,
{\biggl [{5 \over {2\pi G {\rho_{\rm g}}}}\biggr ]^{1/2}},
\end{equation}

\noindent
where $\rho_{\rm p}$ is the bulk density of the planetesimal and $G$ is the gravitational constant. 
ESSTI monitors all these processes until either the planetesimal has left the envelope or 
is fully absorbed (accreted). A planetesimal is considered fully absorbed if it collides with the Jupiter's 
core, breaks up due to ram-pressure, or loses more than 80\% of the mass with which it entered the
envelope, due to ablation.

If the planetesimal leaves the envelope, any mass lost due to ablation is considered accreted and 
is added to the mass of Jupiter. At this stage, ESSTI integrations are stopped and the planetesimal's departing
orbital elements are passed to MERCURY where $N$-body integrations are continued with the planetesimal's
new mass. If the planetesimal returns to the envelope, the above process is repeated with its new mass, new 
radius, and the new state of Jupiter. Integrations are continued until either the planetesimal is fully 
accreted, collided with the Sun or Saturn, or is ejected from the system.

\begin{deluxetable}{ll}
\tablecaption{Planetesimal-Envelope Interaction}
\tablewidth{0pt}
\tablehead{\colhead{\hskip -80pt Effect} & \colhead {\hskip -78pt Formula}} 
\startdata
Gas-drag                      &   $D\pi{R^2}{\rho_{\rm g}}{v_{\rm rel}^2}$                            \\
\\
Aerodynamic Heating           &   ${1\over 4}D {\rho_{\rm g}}{v_{\rm rel}^3}$                         \\
\\
Heating due to ambient gas    &   $\sigma {T_{\rm g}^4}$                                            \\
\\
Radiative Cooling             &   $\sigma {T_{\rm p}^4}$                                            \\
\\
Evaporation Cooling           &   ${P_0}{E_0}{[\pi{m_{\rm p}}/8k{T_{\rm p}}]^{1/2}}{e^{-A/{T_{\rm p}}}}$    \\
\\
Ram-Pressure                  &   ${1\over 2} {\rho_{\rm g}}{v_{\rm rel}^2}$                           \\
\enddata
\end{deluxetable}

\begin{deluxetable*}{cccccccccc}
\tablecaption{Outcome of Integrations (all values are in percentage).}
\tablecolumns{10}
\tablewidth{0pt}
\tabletypesize{\scriptsize}
\tablehead{
\colhead{Planetesimal} & \colhead{Composition} & \colhead{Saturn Mass} & \colhead{Break up} & 
\colhead{Vaporized} & \colhead{Collided} & \colhead{Collided} & \colhead{Ejected} & \colhead{Collided} & 
\colhead{Unaffected}\\
\colhead{Radius (km)} & \colhead{} & \colhead{$(M_{\rm Sat})$} & \colhead{} & \colhead{} & \colhead{Jup. Core} &
\colhead{with Sat.} & \colhead{} & \colhead{with Sun} & \colhead{}
}
\startdata
1   &  Ice   &  1/10  &  67.9  &  1.5  &  0     &  0.2  &  16.2  & 0    &  14.7  \\
1   &  Ice   &  1/3   &  71.7  &  1.0  &  0     &  0.2  &  20.5  & 0    &  6.80  \\
1   &  Ice   &  1     &  75.2  &  0    &  0     &  0.4  &  19.4  & 0    &  4.90  \\
1   &  Rock  &  1/10  &  54.2  &  1.3  &  14.1  &  0.1  &  15.3  & 0    &  14.7  \\
1   &  Rock  &  1/3   &  58.5  &  0.9  &  12.5  &  0.5  &  20.7  & 0    &  6.80  \\
1   &  Rock  &  1     &  51.5  &  0.3  &  19.8  &  2.1  &  21.3  & 0    &  4.90  \\
10  &  Mix   &  1/10  &  68.7  &  0.9  &  0     &  0.1  &  15.4  & 0    &  14.8  \\
10  &  Mix   &  1/3   &  70.6  &  0.8  &  0     &  0.5  &  21.3  & 0    &  6.80  \\
10  &  Mix   &  1     &  71.0  &  0.4  &  0     &  1.5  &  22.0  & 0    &  5.00  \\
100 &  Rock  &  1/10  &  48.5  &  0.1  &  18.0  &  0.4  &  18.3  & 0    &  14.6  \\
100 &  Rock  &  1/3   &  49.4  &  0    &  18.8  &  1.2  &  23.5  & 0.1  &  6.90  \\
100 &  Rock  &  1     &  37.1  &  0    &  23.0  &  4.0  &  30.7  & 0    &  5.20  \\
\enddata
\end{deluxetable*}

\section{Results and Analysis}
We carried out a total of 14,076 four-body integrations corresponding to 1173 integrations for each 
line of size+composition presented in Table 3. In each system, we started Jupiter and Saturn in their current
orbits and randomly chose the semimajor axis of the planetesimal from the range of 
3.7 AU to 6.7 AU (corresponding to 4 Jupiter's Hill radii on either side of the planet). 
This range was chosen so that our initial conditions would be consistent with those in Paper I.
The eccentricity of the planetesimal varied from 0 to 0.05. To avoid computational complications due 
to including Saturn as a growing body, we 
considered the planet in the orbit of Saturn to have been fully formed. However, to examine the dynamical and physical 
consequences of its perturbation, we ran integrations for three different values of its mass corresponding 
to $1/10^{\rm th}$, $1/3^{\rm rd}$ and the full mass of Saturn.
No inclination was considered for planetesimals, and the entire system was considered to be co-planar.

Integrations were carried out for $3\times 10^6$ years, approximately twice the time for the protoplanetary 
envelope to collapse (see figure 1). The $N$-body integrations were carried out using MERCURY's hybrid routine 
with a time step of 10 days and ESSTI integrations were carried out using a $4^{\rm th}$ order Runge-Kutta integrator. 
Trial runs carried out prior to the main inetegrations indicated $10^{-12}$ as the optimal accuracy. We, therefore,
set the accuracy of both integrators to this value. The ESSTI integrator has an adaptive time step capability
that allows for adjusting this quentity for each integration for a given accuracy ($10^{-12}$).

As mentioned before, if a planetesimal entered the envelope, one of the following would occur. Either the
planetesimal would leave the envelope and continue its orbit around the Sun after having lost energy 
and mass due to its interaction with the envelope, or it would continue its motion entirely in the envelope until 
it would collide with Jupiter's core, or it would vaporize completely while inside the envelope, or
it would break up into smaller pieces due to ram-pressure. In the latter case, we assumed that these pieces gradually 
descended toward the center of mass of Jupiter until they either collided with Jupiter's core or fully vaporized.
Table 3 shows the percentage of the occurrence of these outcomes for some combinations of the size and
compositions of planetesimals. In the following, we explain these cases in more detail.

\subsection{Pure-ice planetesimals}
The most notable results shown by Table 3 are those of 1 km pure-ice planetesimals. Because these planetesimals 
are small and volatile, it is natural to expect their motion to be mainly affected by vaporization due 
to heating by gas-drag. However, as shown by columns 5 and 6, it is the ram-pressure that dominates their motion. 
Vaporization is in fact a rare occurrence with no vaporization occurring in systems where Saturn has its full mass, 
and only 1\%-1.5\% of the planetesimals vaporize when the planet in the orbit of Saturn has 1/3 and 1/10 of 
Saturn-mass (figure 2). As shown by columns 4 and 5, when these planetesimals enter the envelope, the majority 
of them breakup at large altitudes (Figure 3) and their small pieces are rapidly (within 
almost one integration time step) vaporized such that none survives to collide with the Jupiter's core intact.

\begin{figure}[ht]
\vskip -15pt
\hskip -24pt
\includegraphics[scale=0.41]{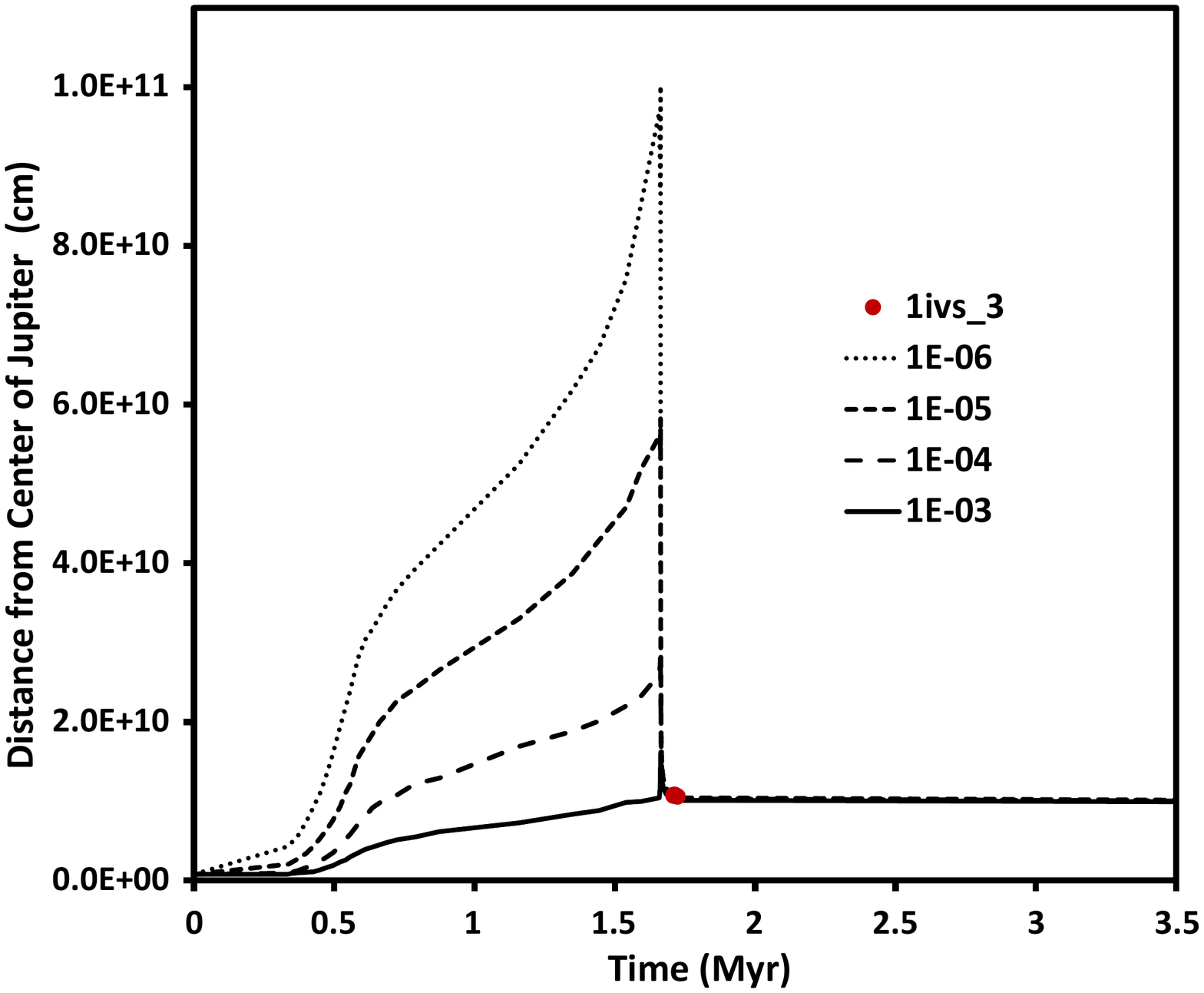}
\vskip -50pt
\hskip -20pt
\includegraphics[scale=0.4105]{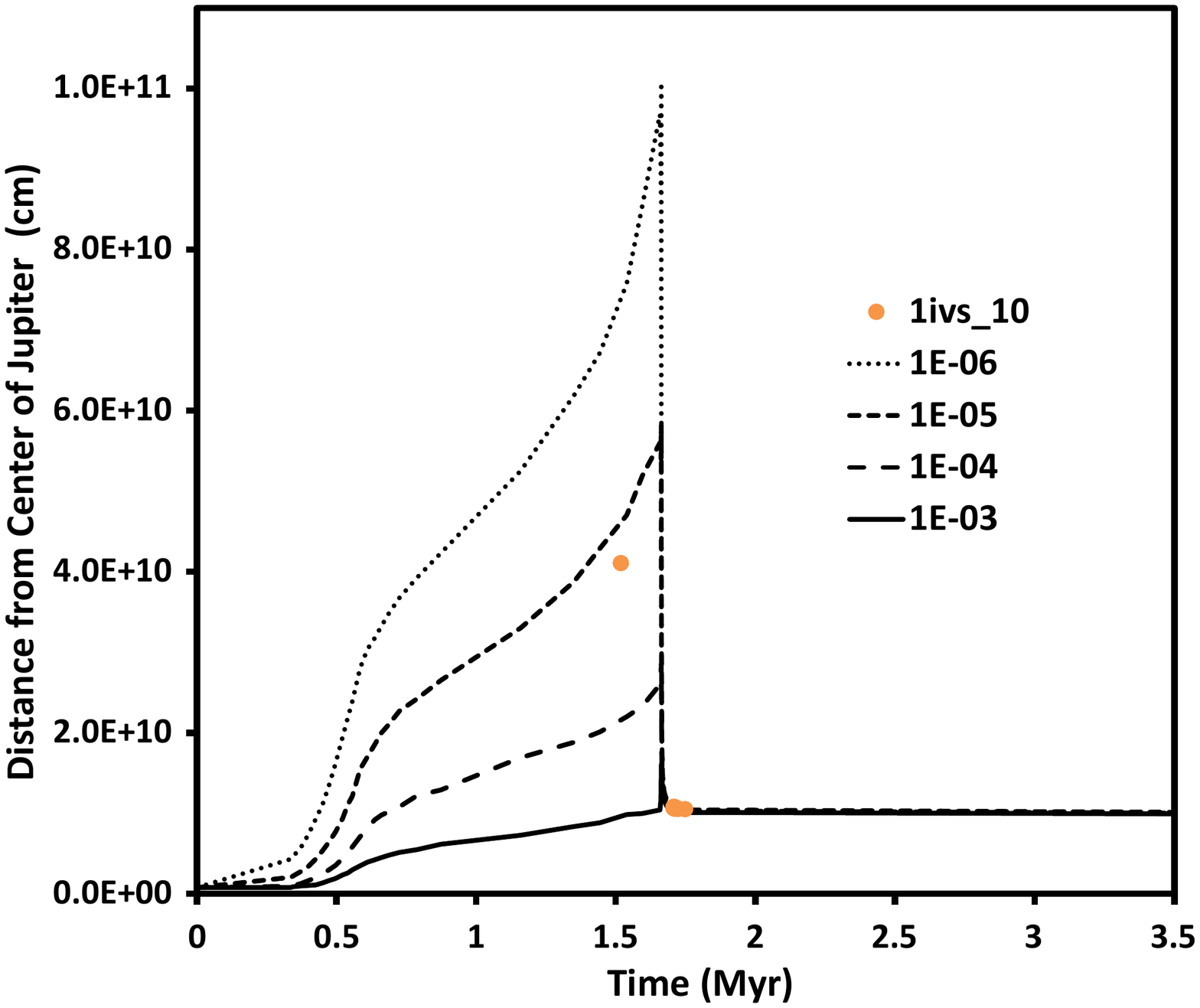}
\vskip -30pt
\caption{Graphs of the vaporization of 1 km pure-ice planetesimals for two values of Saturn's mass. 
Top panel corresponds to 1/3 of Saturn's mass showing only 1\% vaporization, and bottom panel corresponds
to 1/10 of Saturn, showing 1.5\% vaporization. See section 3.1 and Table 3 for more details.}
\label{fig2}
\end{figure}

\begin{figure}[ht]
\vskip -15pt
\hskip -30pt
\includegraphics[scale=0.41]{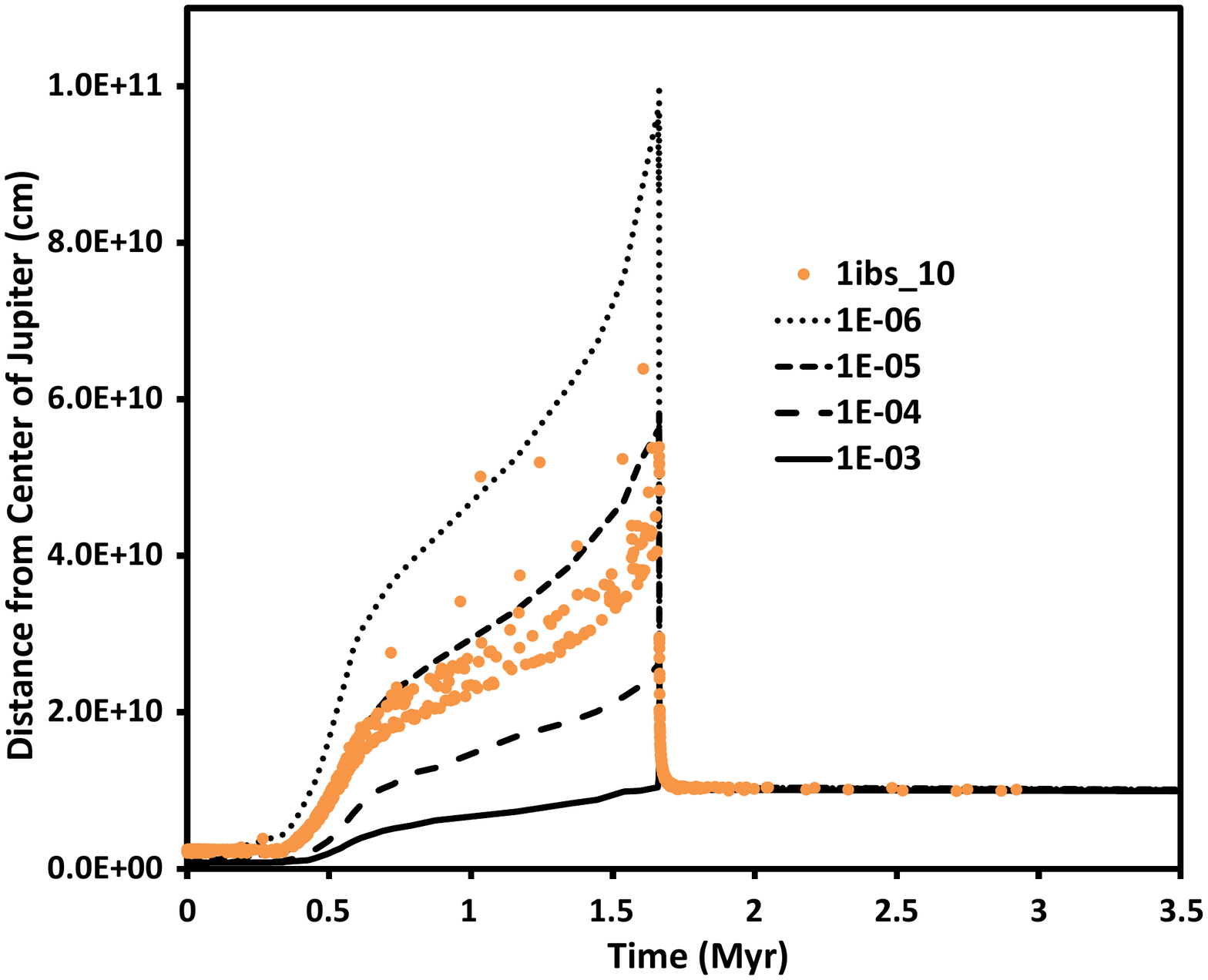}
\vskip -50pt
\hskip -30pt
\includegraphics[scale=0.41]{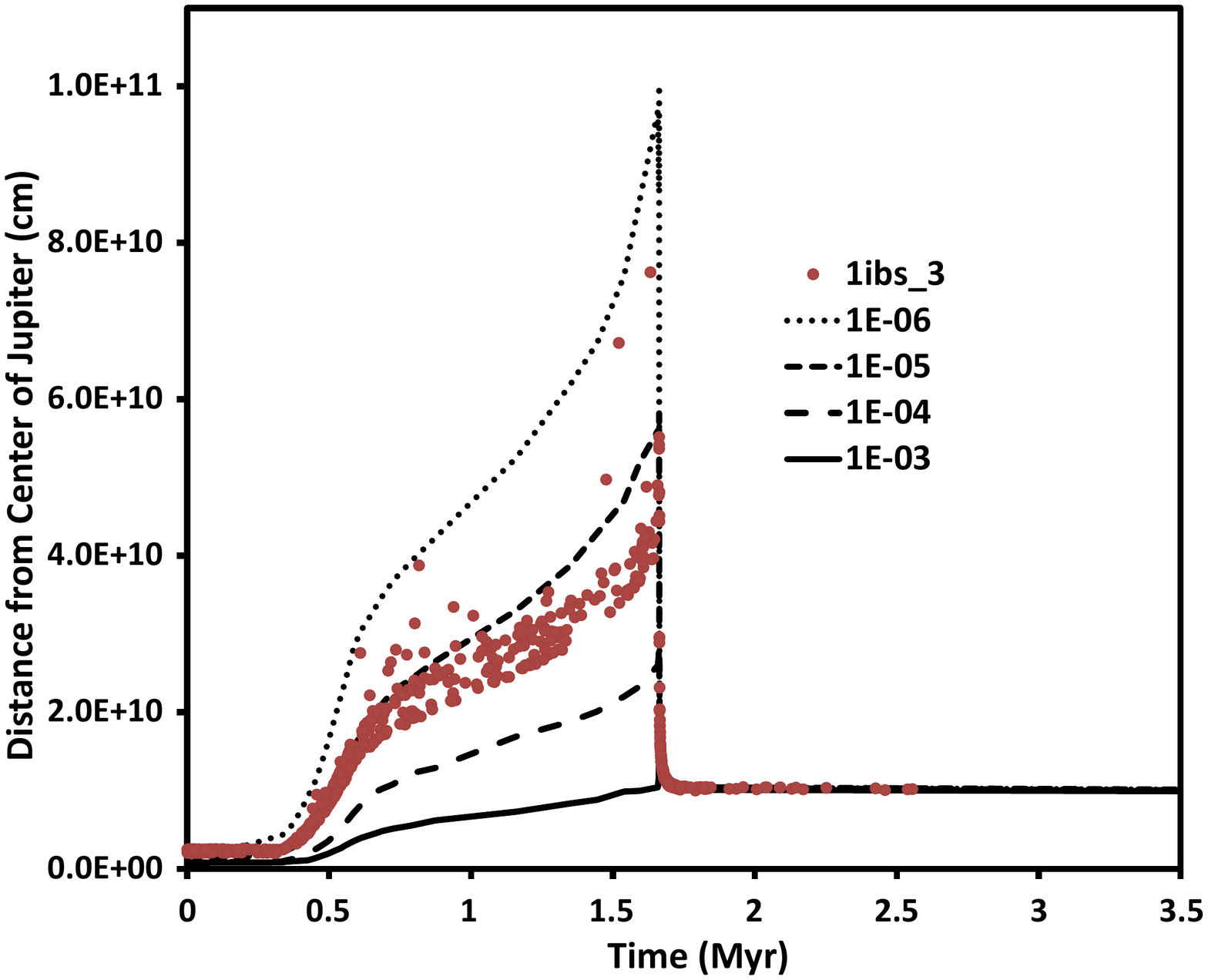}
\vskip -50pt
\hskip -30pt
\includegraphics[scale=0.41]{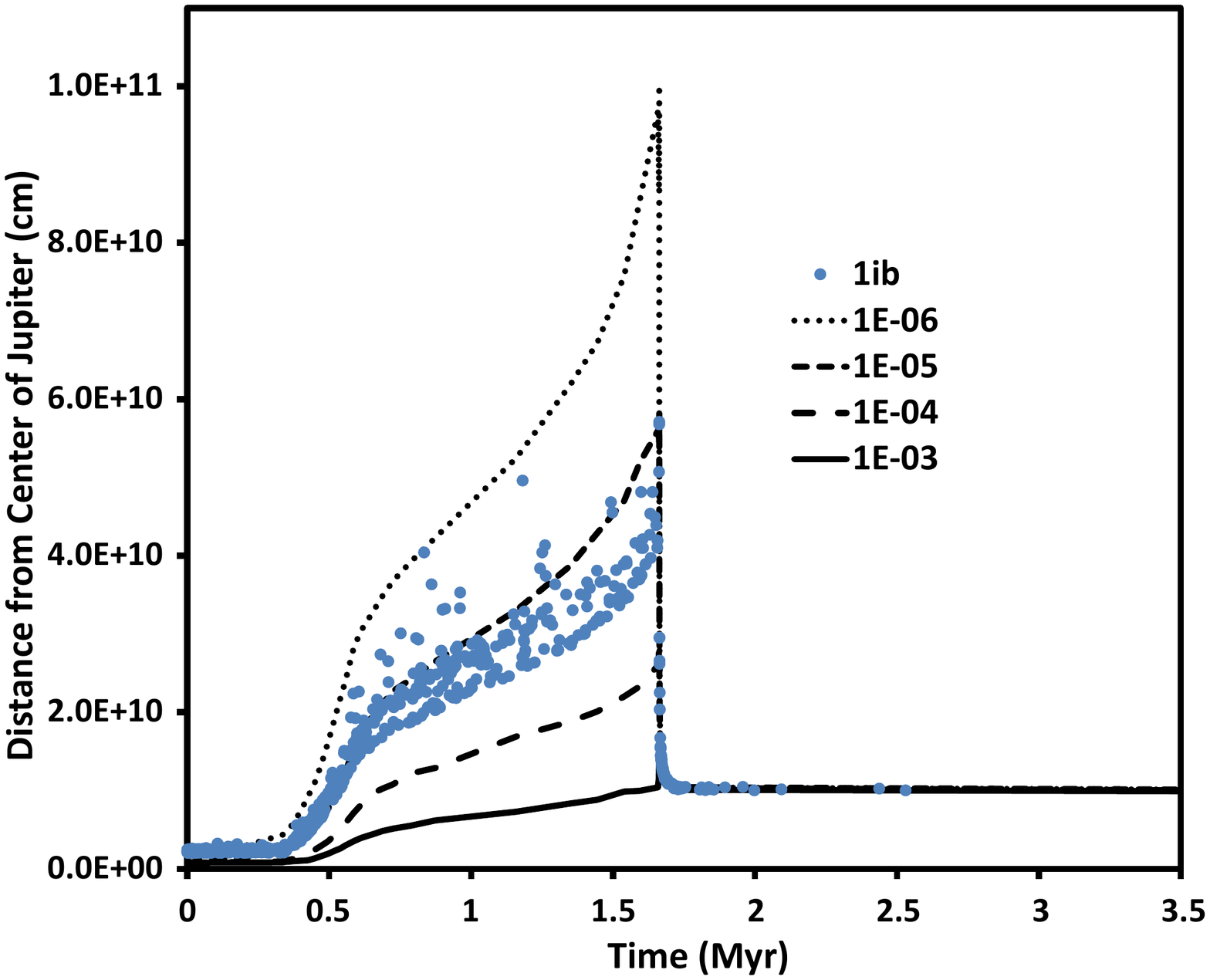}
\vskip -20pt
\caption{Graphs of the break-up of 1 km pure-ice planetesimals due to ram-pressure for different
values of Saturn's mass. From top to botom, panels corresponds to $1/10^{\rm th}$, $1/3^{\rm rd}$, and 
full Saturn mass. As shown here, break up happens at high altitudes in the envelope.}
\label{fig3}
\end{figure}

\begin{figure}[ht]
\vskip -15pt
\hskip -25pt
\includegraphics[scale=0.41]{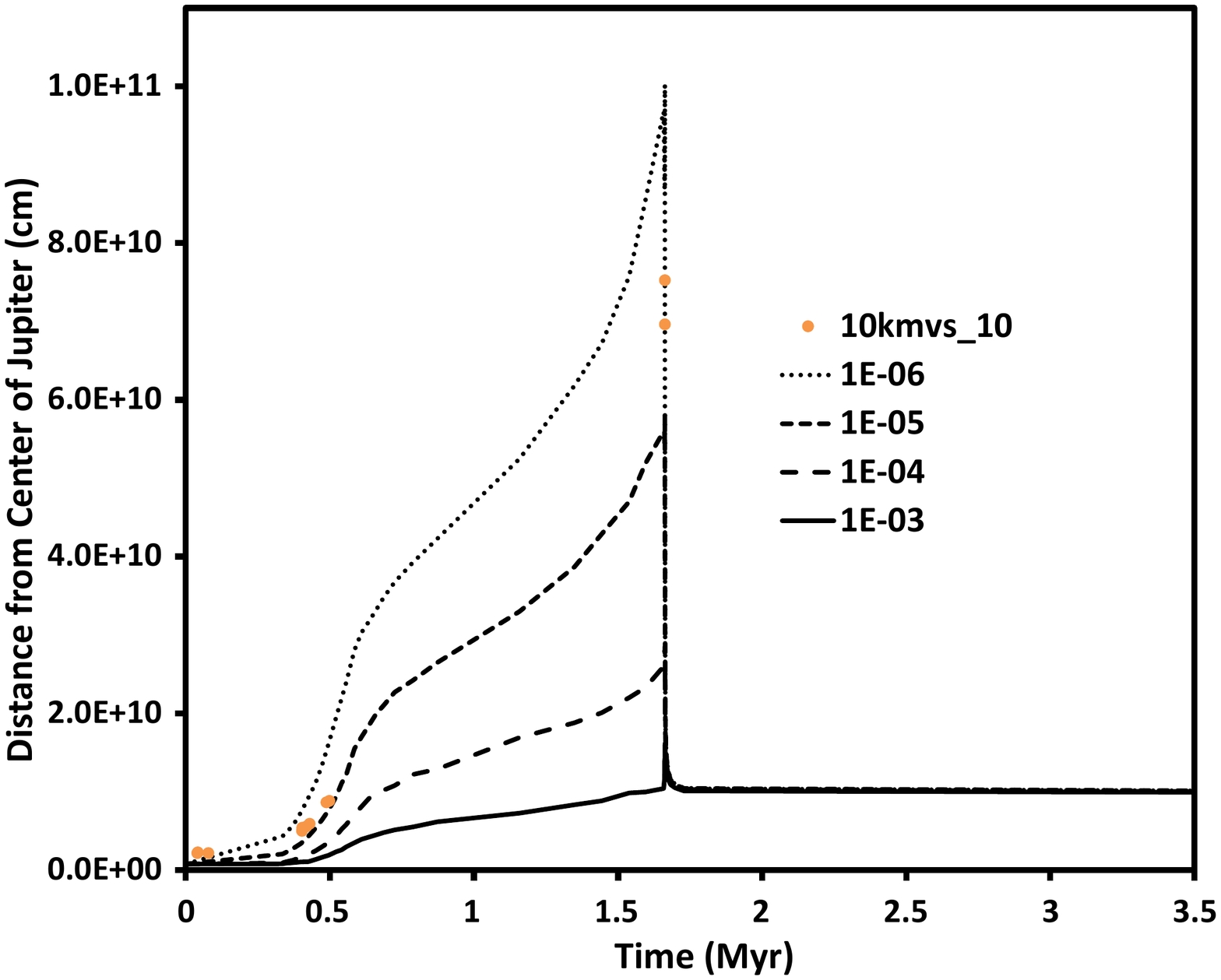}
\vskip -50pt
\hskip -40pt
\includegraphics[scale=0.41]{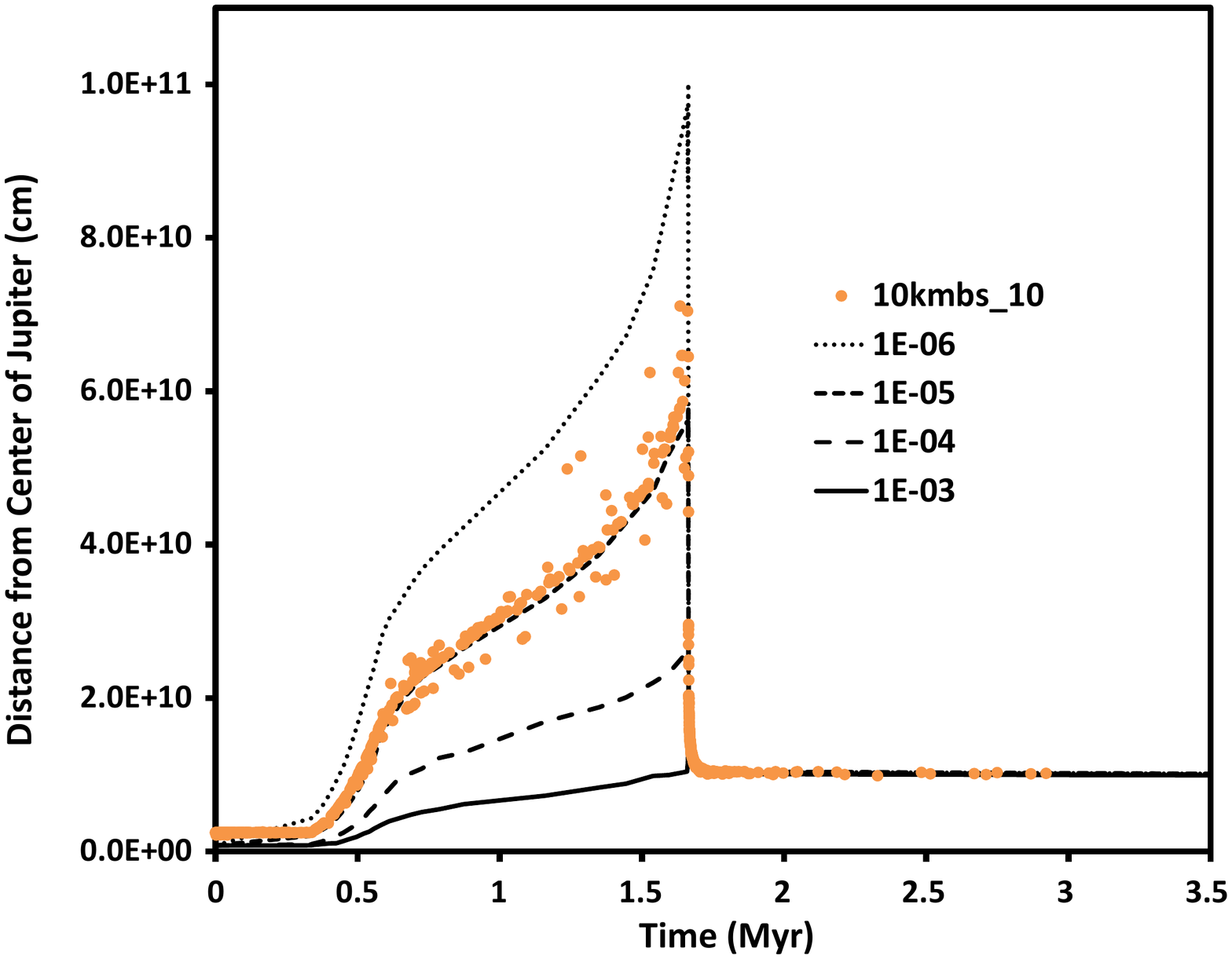}
\vskip -30pt
\caption{Graphs of the vaporization (top) and break-up (bottome) of 10 km mix-composition planetesimals
for $1/10^{\rm th}$ of Saturn mass. As shown here, vaporization is rare and ram-pressure break-up is the dominent
process.}
\label{fig4}
\end{figure}

\subsection{Mixture of ice and rock}
As mentioned earlier, when a mix-composition planetesimal is heated, its ice component vaporizes faster and
carries some of its rocky grains with it. This rocky debris is then released into the envelope and enhances the
envelope's metalicity. Because in this case, the planetesimal loses mass in the form of both ice and dust, 
the rate of its mass-loss is higher than other planetesimal types. Although
this higher rate of mass-loss causes the planetesimal's surface-to-mass ratio to increase, which in turn makes
the planetesimal more susceptible to be fully accreted through heating by the gas-drag, its larger bulk density 
$(2 \, {\rm g \, cm^{-3}})$ gives it more inertia making its rocky component more resistive to vaporization 
and breakage due to the ram-pressure. The 10 km mix-composition planetesimals, as intermediates between the 
1 km and 100 km bodies, offer the best case to examine how these two competing 
effects play off against each other. Figure 4 shows this in more detail for $1/10^{\rm th}$ of Saturn mass.
The top panel in this figure shows the 
time and location of vaporization, and the bottom panel shows the same quantities for breakage due to the ram-pressure. 
As shown here, similar to the case of pure-ice planetesimals, ram-pressure has the dominant effect causing the 
mix-composition planetesimal to break up relatively high in the envelope. The vaporization, on the other hand, 
is rare and occurs mainly at the early stages of Jupiter's evolution.

\begin{figure}[ht]
\vskip -6pt
\hskip -30pt
\includegraphics[scale=0.41]{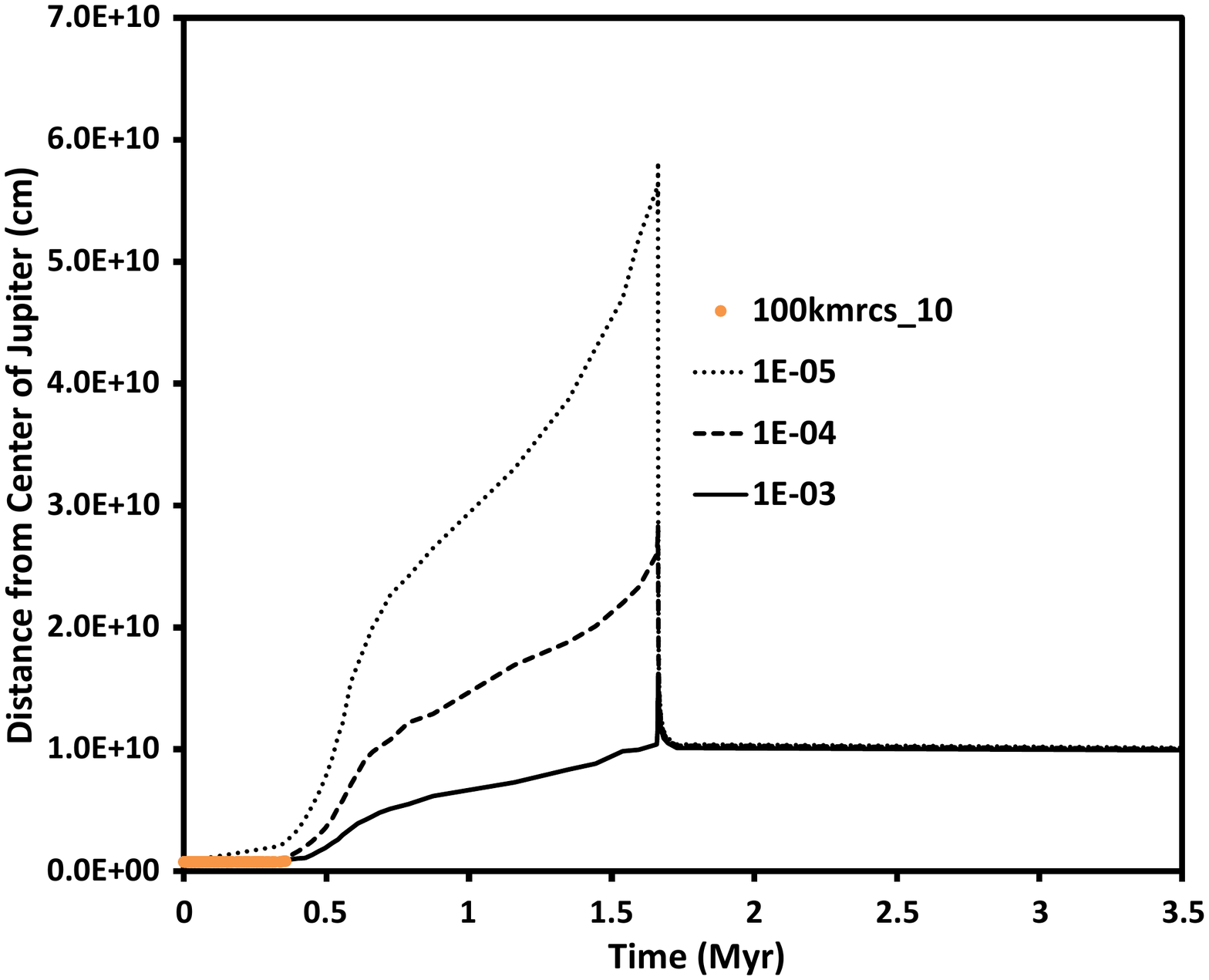}
\vskip -50pt
\hskip -30pt
\includegraphics[scale=0.41]{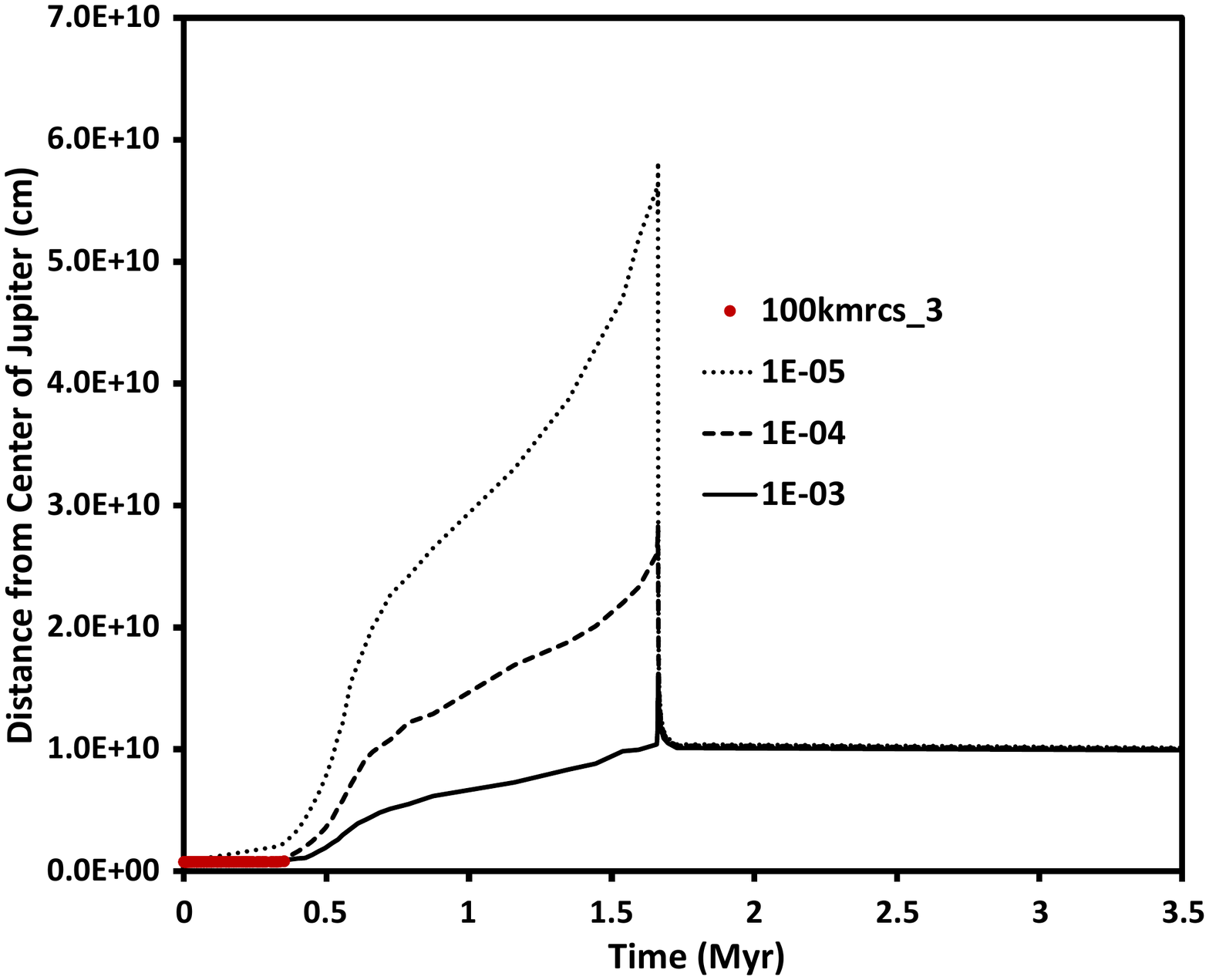}
\vskip -50pt
\hskip -30pt
\includegraphics[scale=0.41]{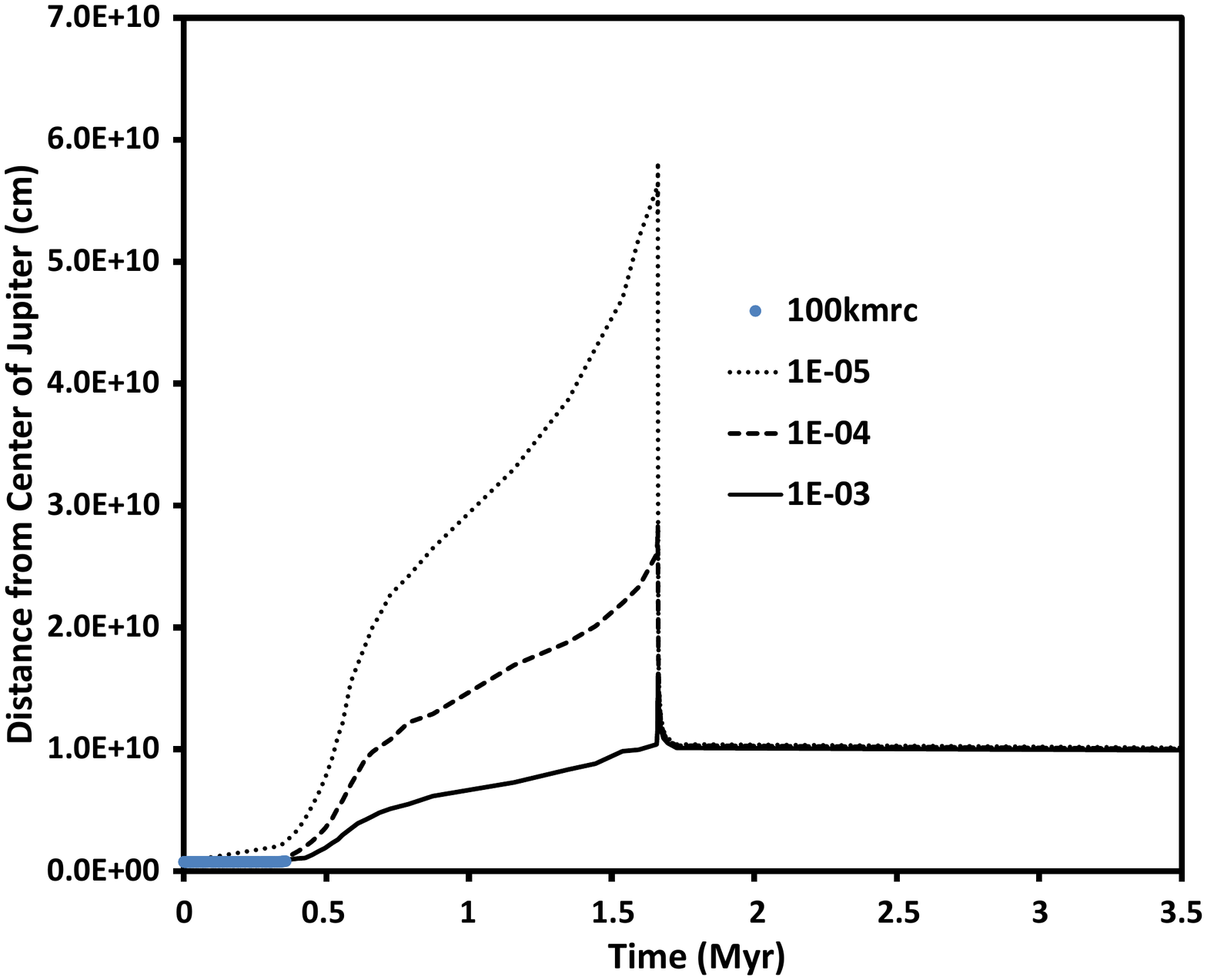}
\vskip -27pt
\caption{Graphs of 100 km rocky planetesimals penetrating the envelope and striking the surface of Jupiter's core
at early times. From top to bottom, the panels correspond to $1/10^{\rm th}$, $1/3^{\rm rd}$, and full mass of Saturn.}
\label{fig5}
\end{figure}

\begin{figure}[ht]
\vskip -6pt
\hskip -30pt
\includegraphics[scale=0.41]{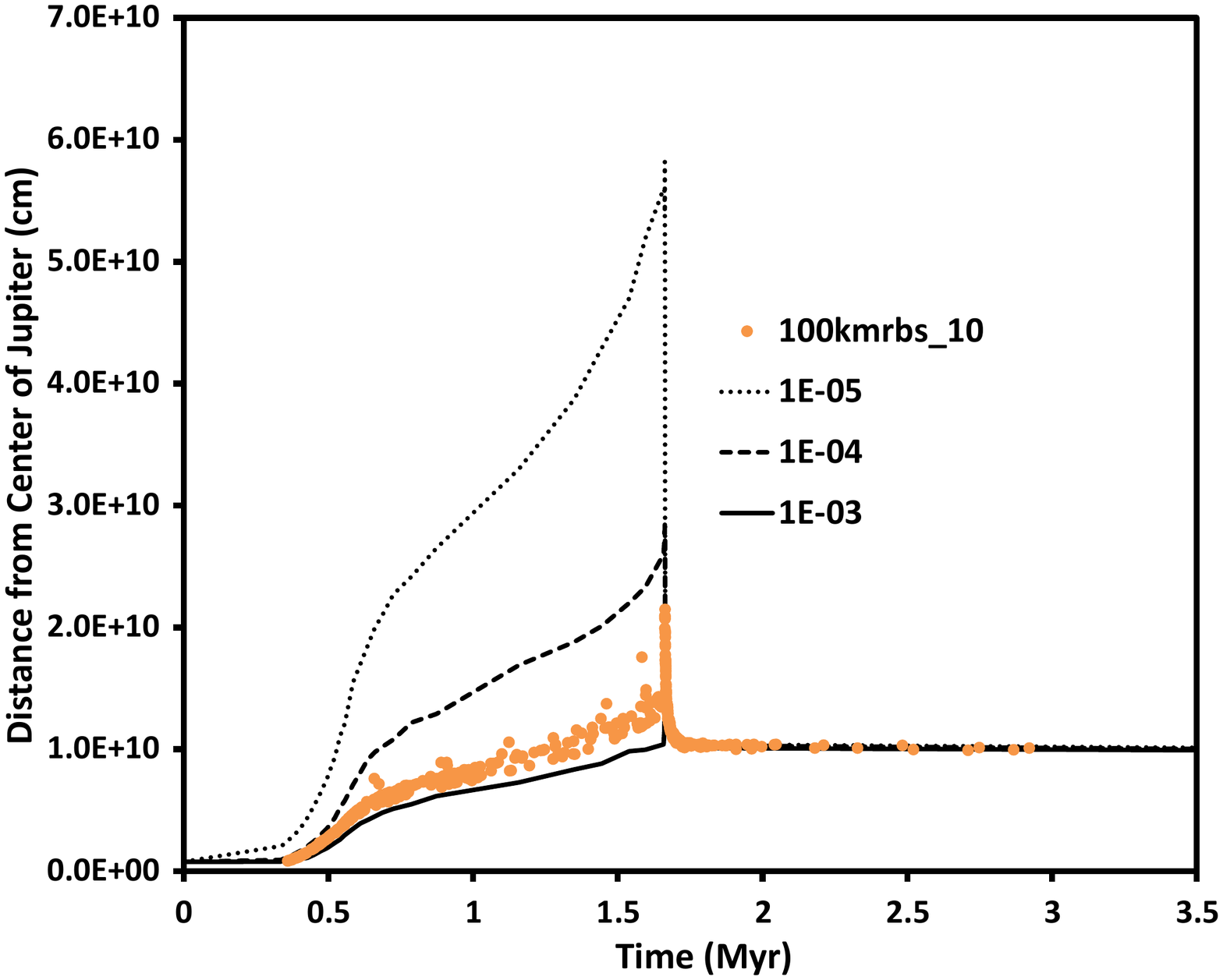}
\vskip -50pt
\hskip -30pt
\includegraphics[scale=0.41]{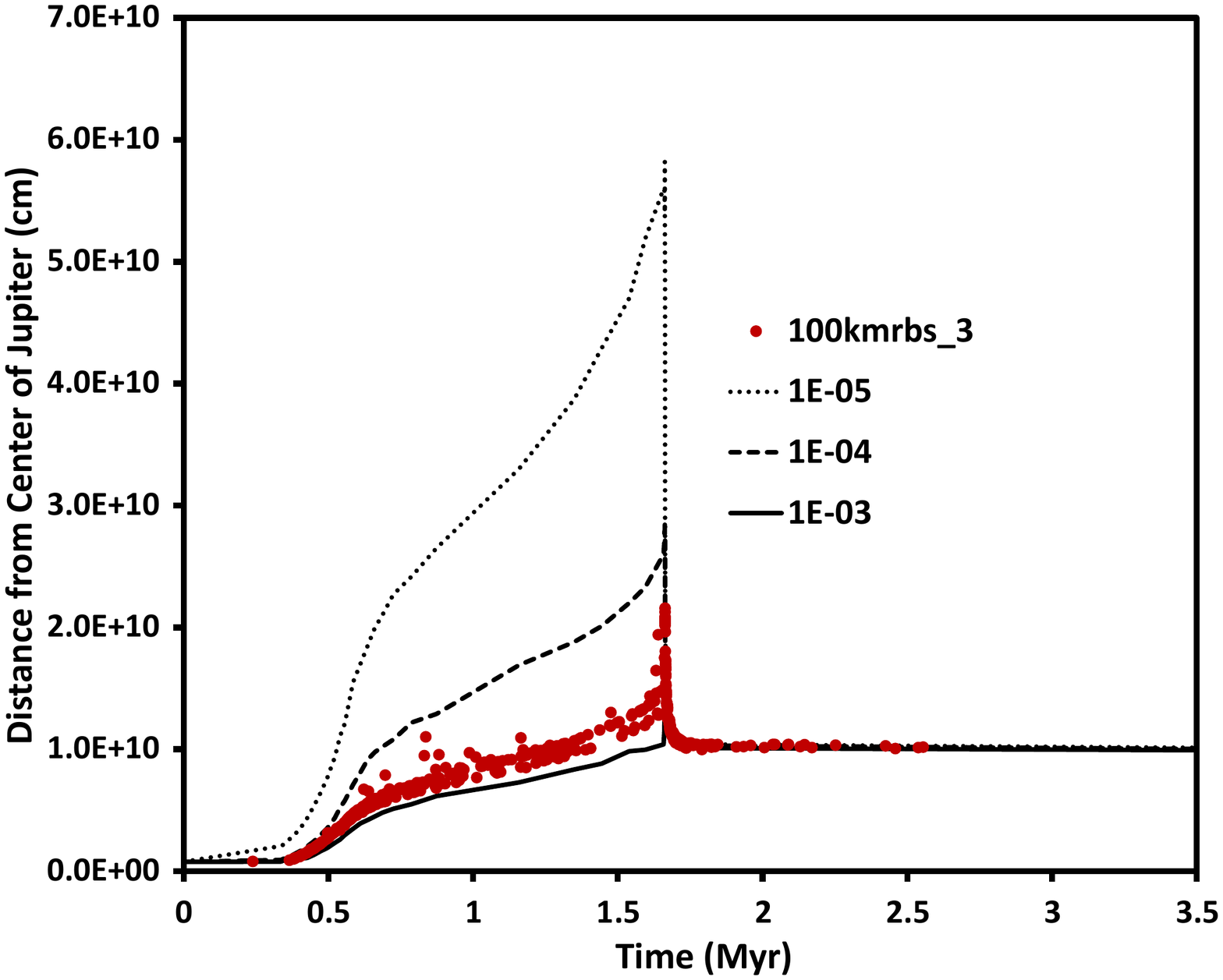}
\vskip -50pt
\hskip -30pt
\includegraphics[scale=0.41]{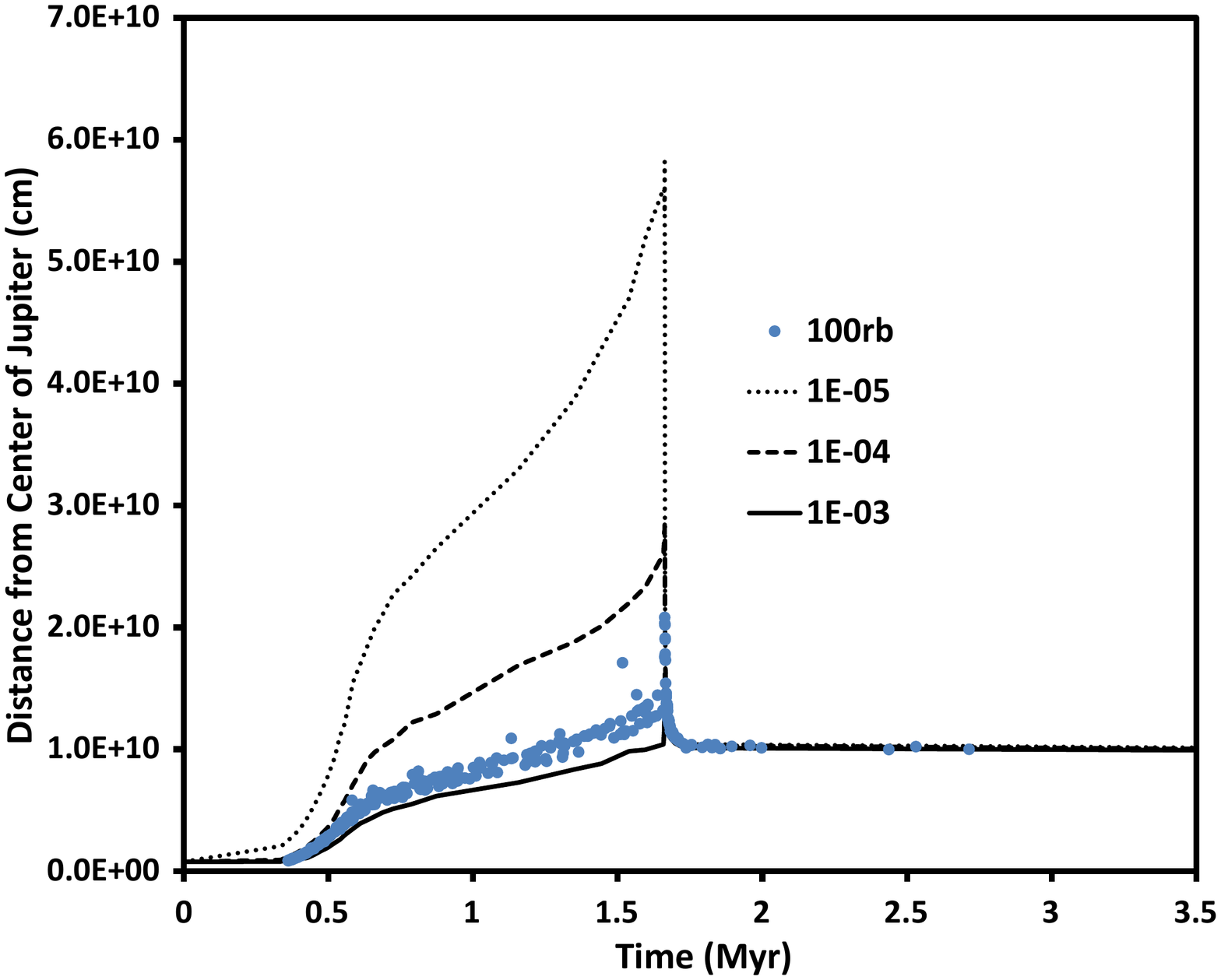}
\vskip -28pt
\caption{Graphs of 100 km rocky planetesimals breaking up by the ram -pressure before their fragments collide with the 
Jupiter's core. From top to bottom the pansl correspond to $1/10^{\rm th}$, $1/3^{\rm rd}$, and full mass of Saturn.
As shown here, break-up occurs deep in the envelope.}
\label{fig6}
\end{figure}

\begin{figure*}[ht]
\center
\vskip -5pt
\hskip -35pt
\includegraphics[scale=0.5]{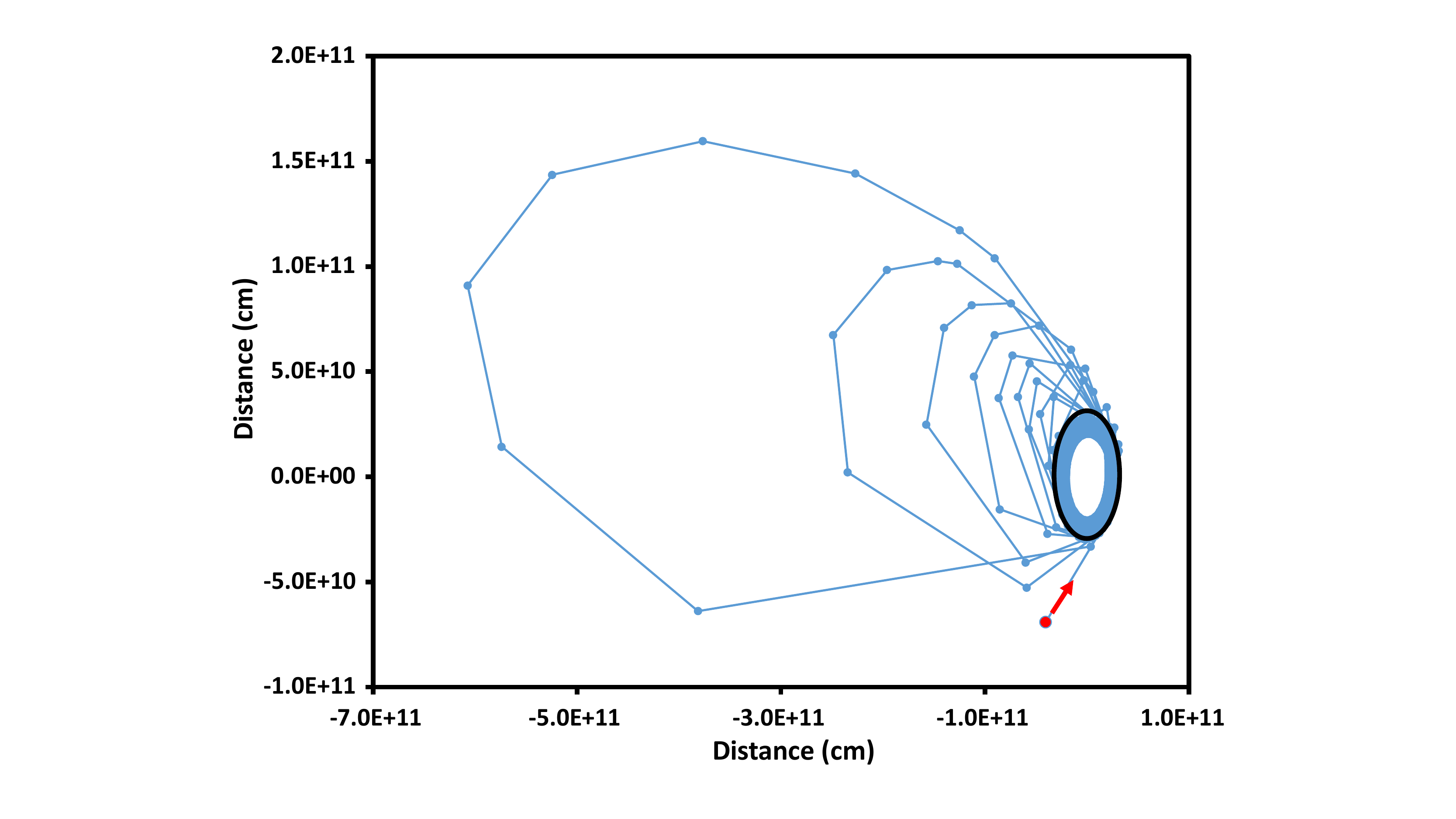}
\vskip -20pt
\hskip -20pt
\includegraphics[scale=0.47]{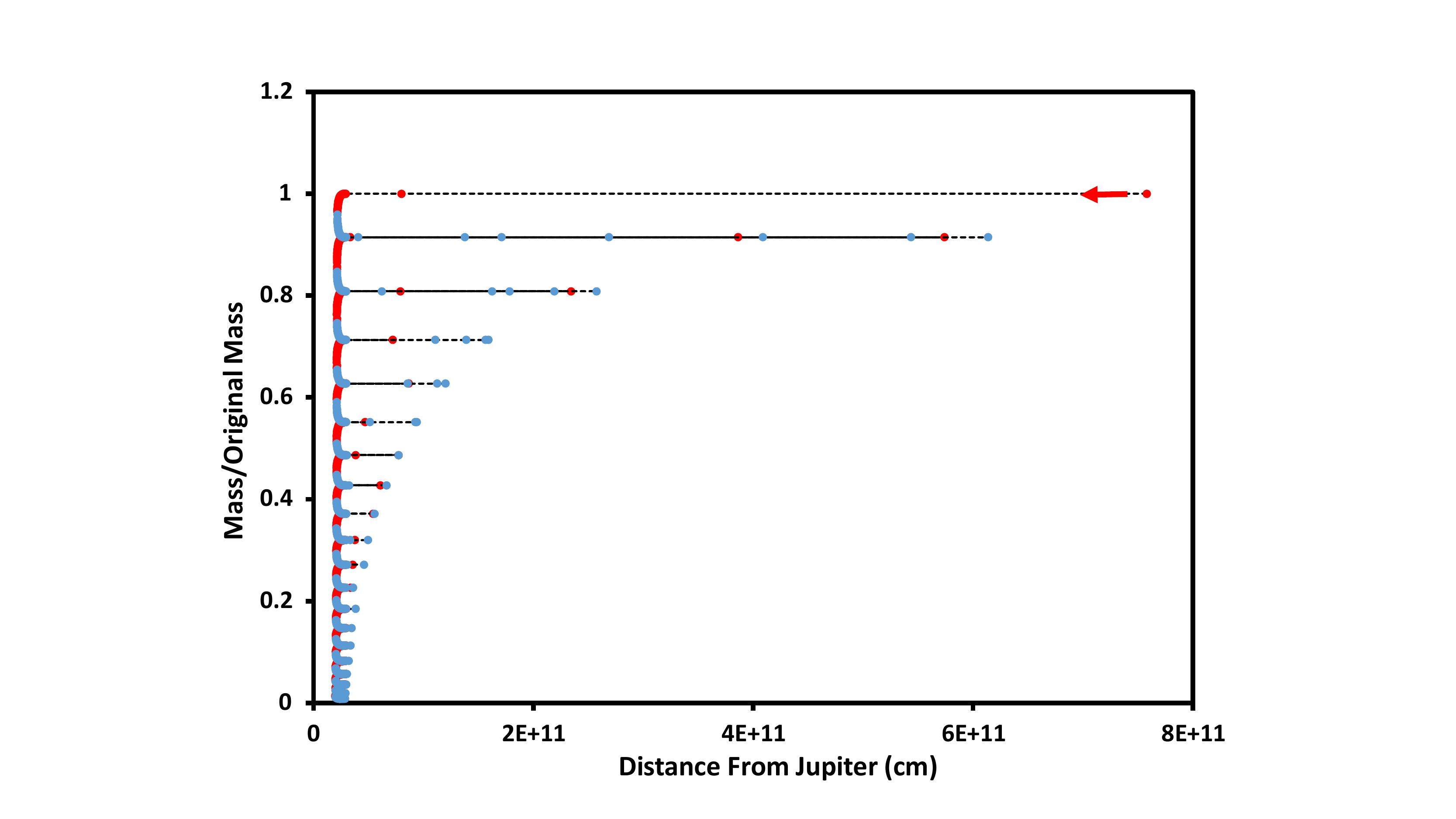}
\vskip -10pt
\caption{Interaction of a 100 km rocky planetesimal with the envelope. The top panel shows the planetesimal's
actual path. The dark oval shows the outer edge of the envelope. The position and motion of the planetesimals prior 
to encountering the envelope is shown by the red arrow. Red points indicate inward motion towards the core, and blue 
points indicate outward motion away from the core. As shown here, the planetesimal enters and leaves the envelope 
several times until it is fully captured. The bottom panel shows the amount of the mass that planetesimal loses during
each passage through the envelope.}
\label{fig7}
\end{figure*}

\begin{figure*}[ht]
\center
\vskip -5pt
\hskip -35pt
\includegraphics[scale=0.5]{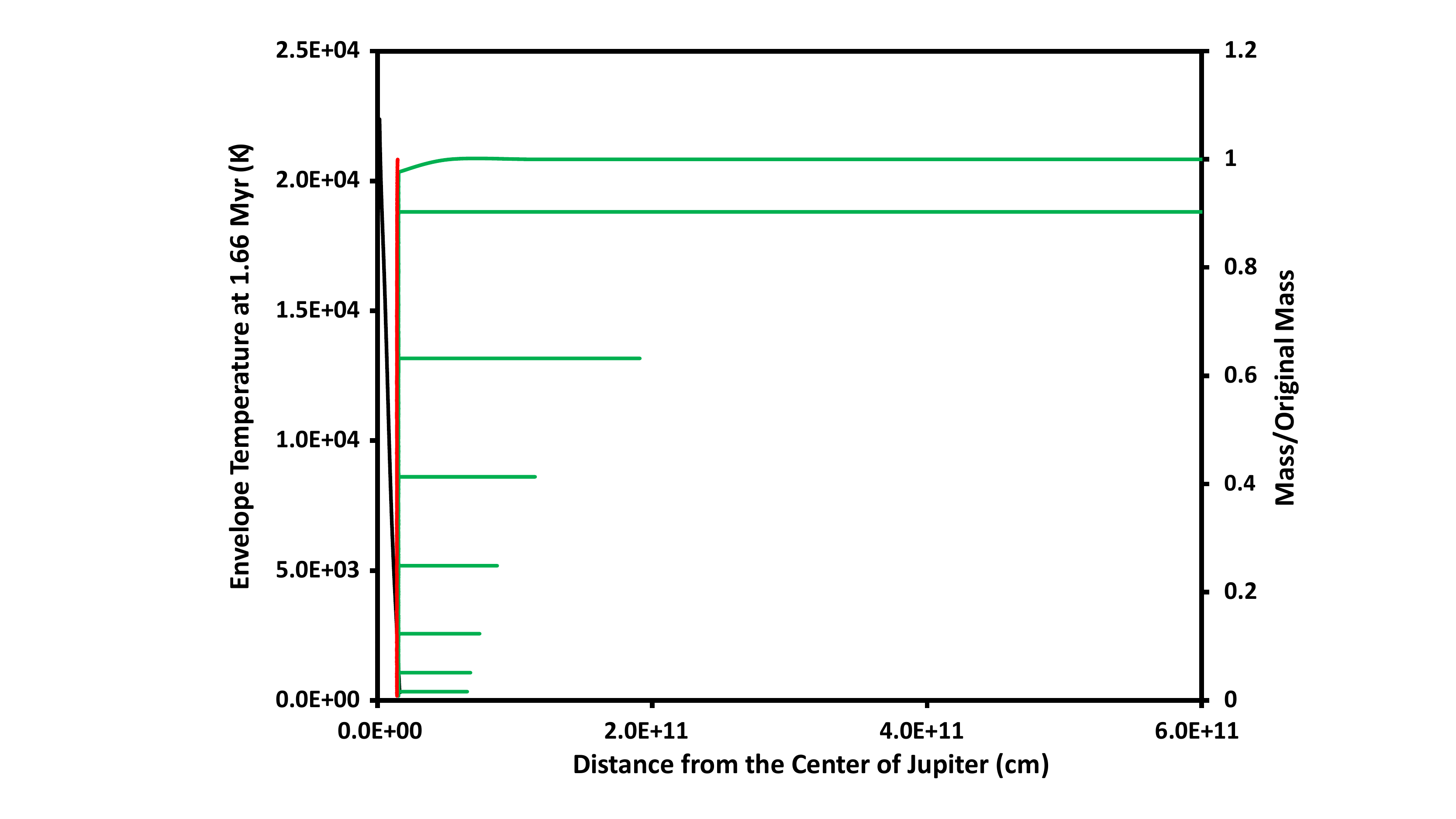}
\vskip -5pt
\caption{Graphs of the mass-deposition of two 1 km rocky planetesimals in term of distance from the center of Jupiter.
Note that the left vertical axis shows the envelope's temperature and the right vertical axis shows the planetesumal's
relative mass. The black line represents the temperature of the envelope at 1.66 Myr. The red curve shows a planetesimal
that penetrates through the envelope and is fully absorbed. The green lines show a planetesimal which undergoes multiple
encounters until it is fully accreted. Note that in both cases, full accretion occurs at $\sim 3000$k.}
\label{fig8}
\end{figure*}

\subsection{Pure-rock planetesimals}
As expected, rocky planetesimals show strong resistance to vaporization. For instance, as shown by figure 5,
$25\%-30\%$ of 100 km pure-rock bodies that encountered the envelope, penetrated through and impacted the surface of 
Jupiter's core at early times and without losing much of their masses. However, and despite the latter, 
ram-pressure was still the dominent factor affecting the motion of these bodies. 
While this seems to be similar to the situation with icy and mix-composition planetesimals, it differs from those cases
in the sense that the higher material strength of rock does not allow the ram-pressure to break up 
rocky planetesimals as easily. Figure 6 shows this for the 100 km rocky planetesimals that encountered the envelope.
As shown here, break-ups occur later and in deeper layers. This suggestes that pure-rock planetesimals
may pass through the envelope several times and lose mass until their self-gravity becomes so weak that
it cannot hold them together against the ram-pressure. Figure 7 shows this for a 100 km rocky planetesimal.
The top panel in this figure shows the actual path of the body and the bottom panel shows the variations in its mass as 
a function of its distance from the center of mass of Jupiter. The black oval in the top panel shows the outer boundary
of the envelope extending to approximately $3 \times 10^{10}$ cm from Jupiter's center of mass. The vertical axis in 
the bottom panel shows the instantaneous mass of the planetesimal relative to its initial mass $(M_0)$. 
The red arrows on both panels show the location of the planetesimal and the direction of its motion.
When approaching the envelope, the planetesimal is shown in red and the variations in its mass are 
presented by dashed lines. When moving away from the envelope, the planetesimal is in blue and its mass is 
shown by solid lines. 

As shown here, at $t=0$, the planetesimals is at approximately $7.5 \times {10^8}$ cm from the center of Jupiter
(the dotted line represening $M/{M_0}=1$ in the bottom panel). When it starts moving toward the envelope, it maintains
its mass until it passes a critical distance at about $2 \times {10^{10}}$ cm where it 
starts to lose mass due to evaporation. The ambient temperature at this region is approximately 2700 K
meaning that, in addition to the gas drag, the hot envelope also contributes to the heating leading to mass loss.
Although the planetesimal has lost mass and therefore, kinetic energy, it still retains enough energy 
to leave the envelope. However, its new kinetic energy is not large enough for escaping Jupiter's Hill sphere completely.
As a result, the planetesimal returns to the envelope and repeats this process while losing energy and mass in each 
encounter until it can no longer leave the envelope. At this stage, either it breaks apart with some of its fragments 
hitting the surface of Jupiter's core (see column 4 of Table 3) or it is fully vaporized.  As can be seen from the figure, 
the vaporization occurs almost entirely in the region where the temperature is above $\sim$2700K.

\subsubsection{The case of 1 km pure-rock planetesimals}
As shown by Table 3, similar to the case of 100 km rocky bodies, 1 km pure-rock planetesimals, too, show large 
number of impacts with Jupiter's core. The main difference, however, is that because of the small sizes of these bodies, 
their motion is more complex. On the one hand, rock is much less volatile than ice, implying that a 1 km rocky planetesimal 
can survive longer in the envelope than a 1 km ice or mix-composition object. On the other hand, because of its small 
radius, the effects of gas-drag can be more significant causing the planetesimal to be more easily absorbed. Figure 8 shows 
the outcome of the interaction of two of such bodies with the envelope.
The vertical axis on the left shows the temperature of the envelope and the one on the right shows the instantaneous and relative
mass of the planetesimal. The black curve shows the envelope's temperature as a function of distance from the center 
of Jupiter at the time 1.66 Myr. The red curve shows an example of those planetesimals whose trajectories take them directly 
into the envelope. In the case of this specific planetesimal, almost the entire mass-loss occurs in the region  
where the temperature is $\sim 3000$ K. The green curve shows an example of those planetesimals that
have grazing encounters with the envelope. In this case, the planetesimal shows similar behavior as the 100 km body:
While it loses some mass and energy, it cannot escape Jupiter’s Hill sphere and makes multiple returns until it is entirely 
absorbed. In both cases, the vaporized mass is deposited in the same region of the envelope that it is produced. 
It should be noted that although the deposition of these materials is in a very limited region, convective mixing is 
likely to spread them over a significantly larger region of the envelope. It is also important to emphasize that
the scenarios presented in figures 7 and 8 can be affected by the presence of the nebular gas outside of 
the envelope especially that these effects are mass (and composition) dependent.

\begin{figure*}[ht]
\hskip -18pt
\includegraphics[scale=0.41]{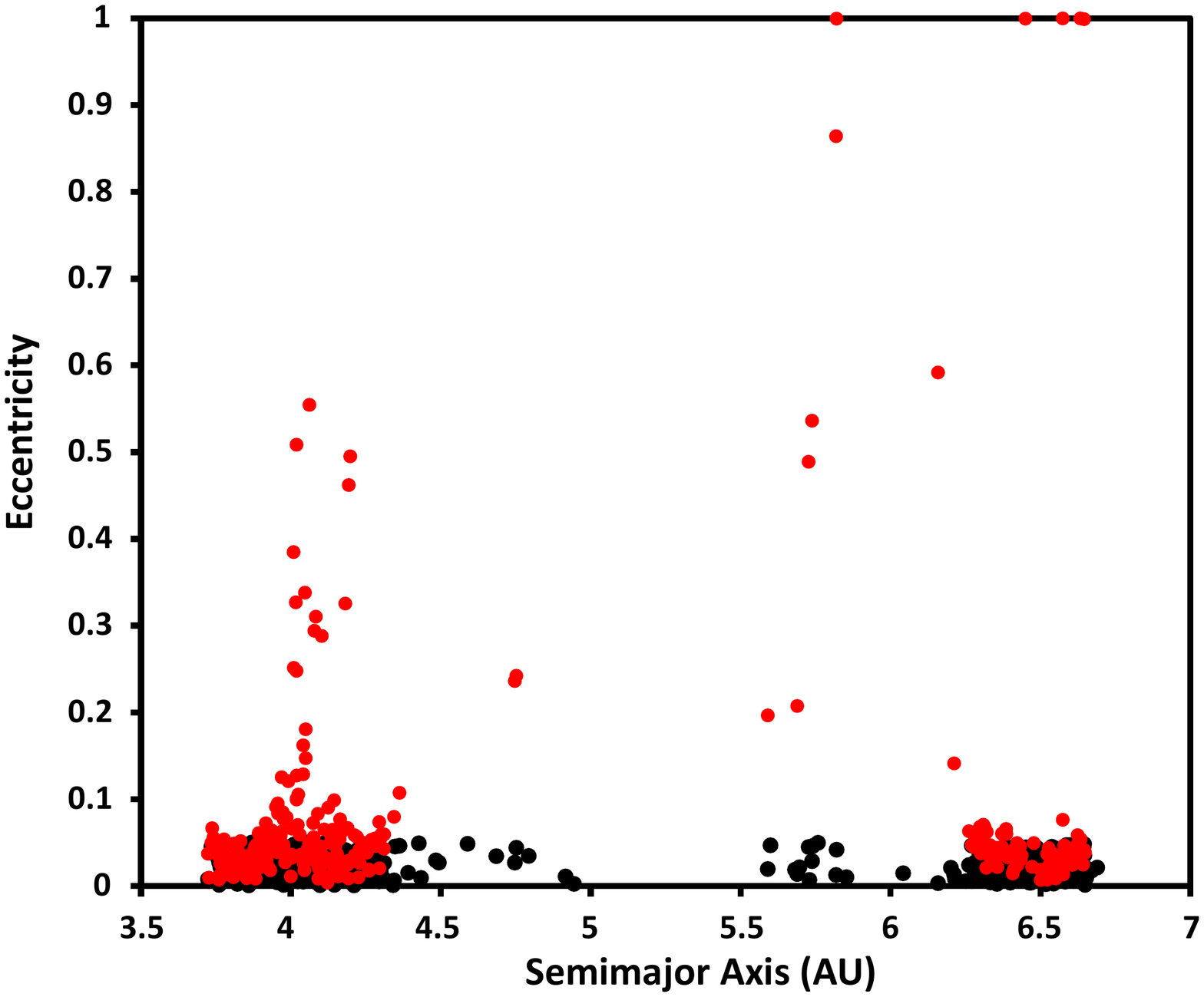}
\hskip -80pt
\includegraphics[scale=0.41]{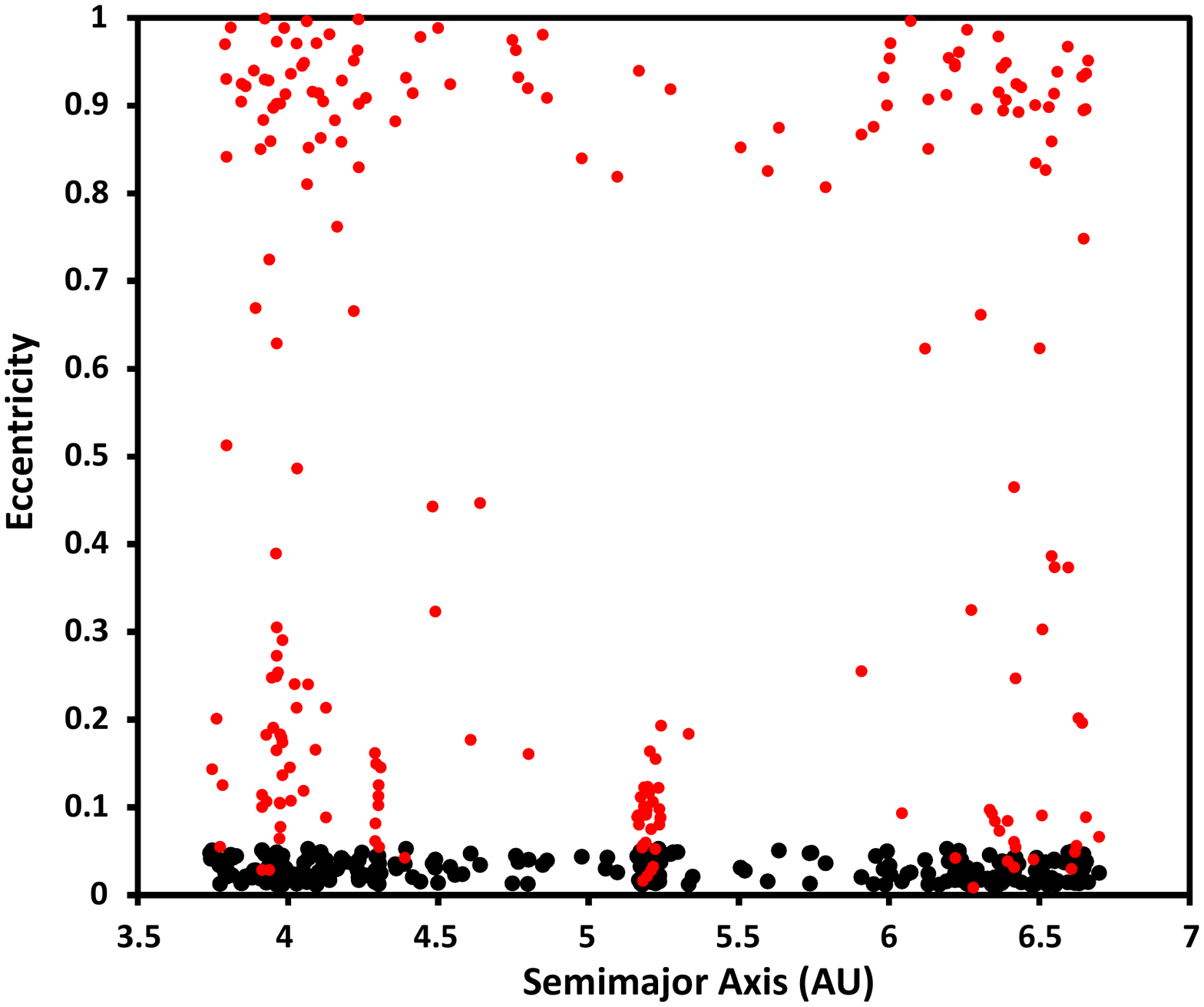}
\vskip -40pt
\hskip -18pt
\includegraphics[scale=0.41]{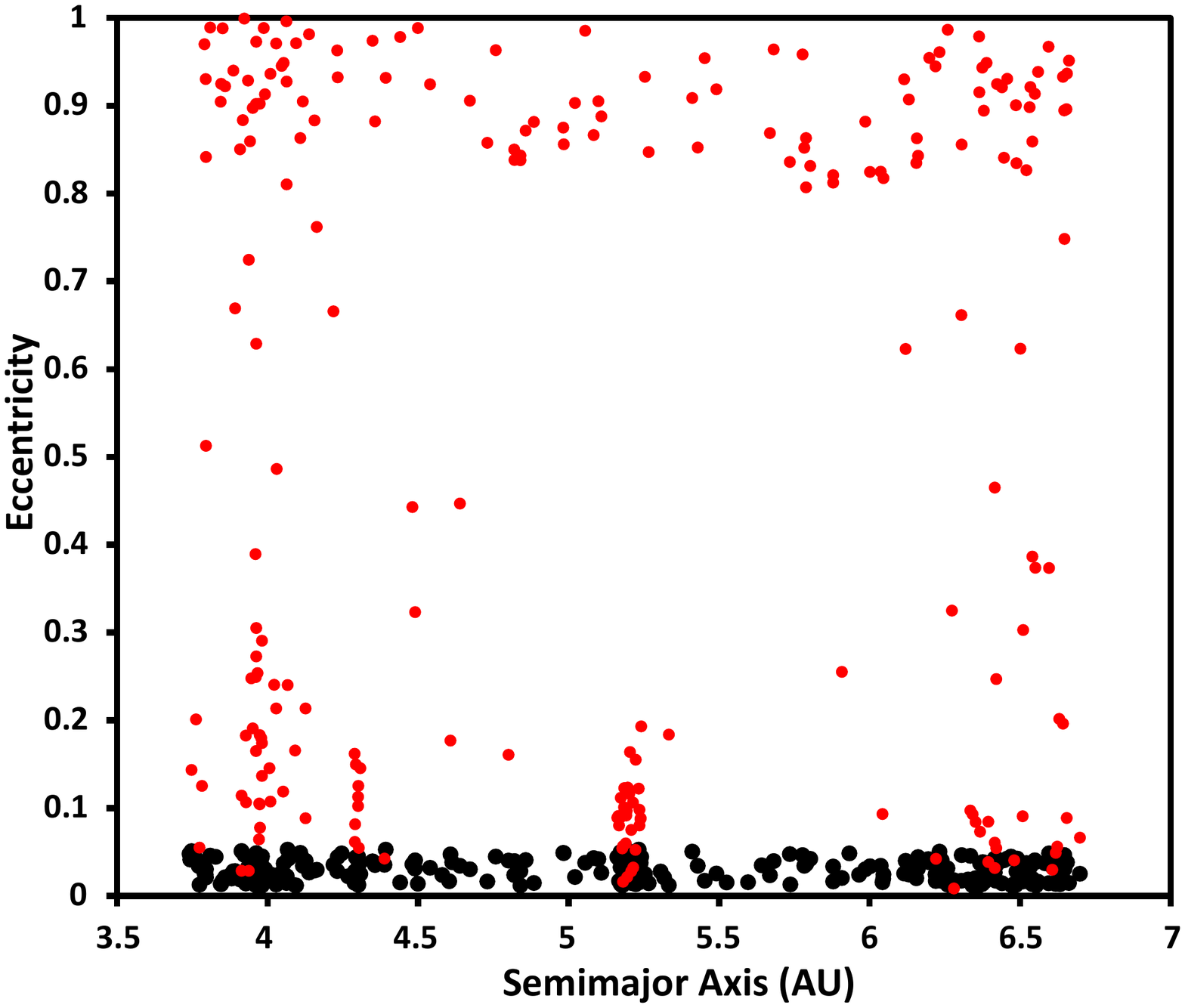}
\hskip -80pt
\includegraphics[scale=0.41]{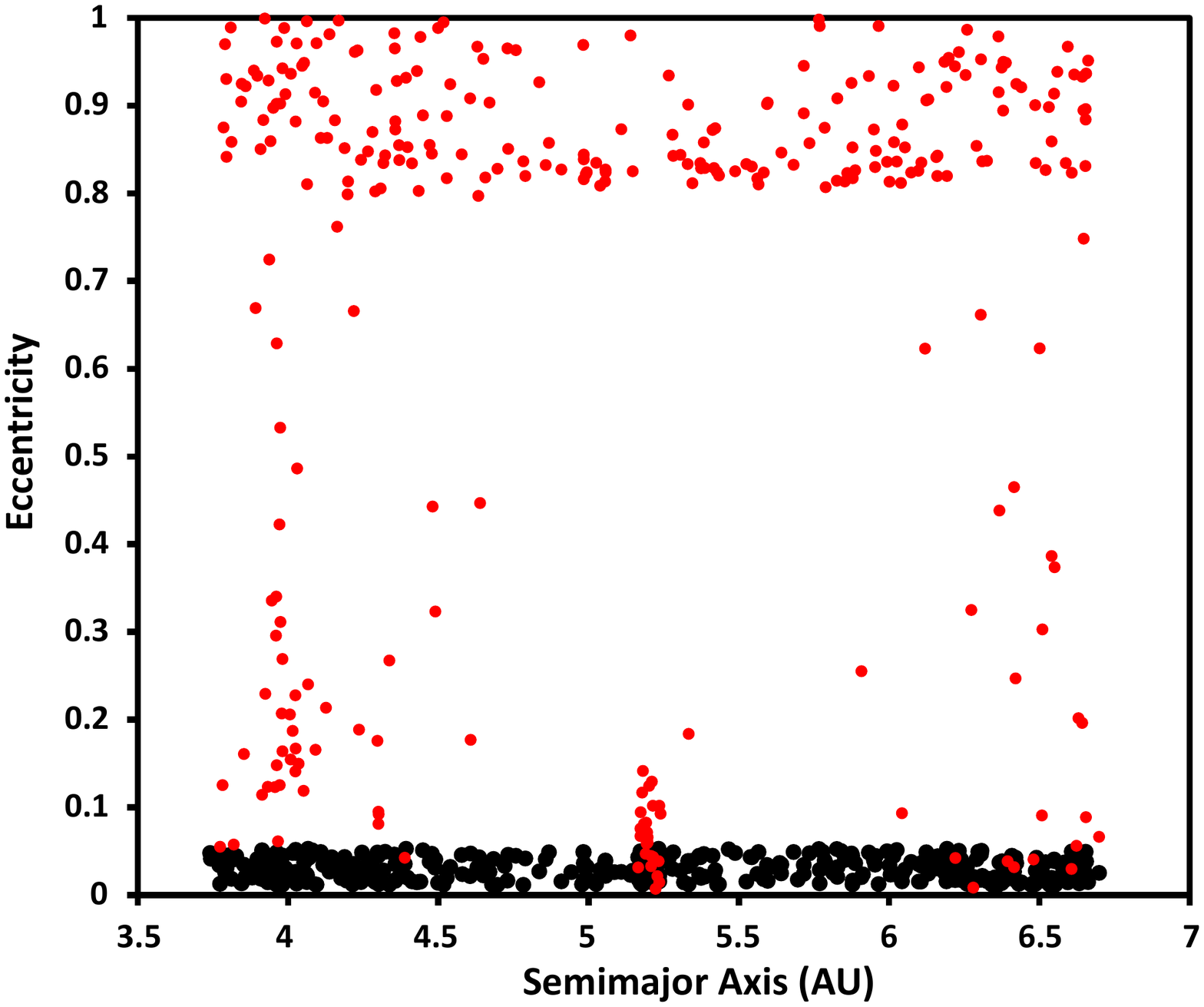}
\vskip -20pt
\caption{Graphs of the final orbital eccentricities of 100 km rocky planetesimals in terms of their initial semimajor
axes. The top-left panel shows the results for the system without Saturn. The top-right panel corresponds to 1/10
of Saturn's mass. Bottom-left panel is for the system with 1/3 of the mass of Saturn, and the bottom-right panel
shows the results for the system with full Saturn-mass.}
\label{fig9}
\end{figure*}

\begin{figure*}[ht]
\hskip -21pt
\includegraphics[scale=0.38]{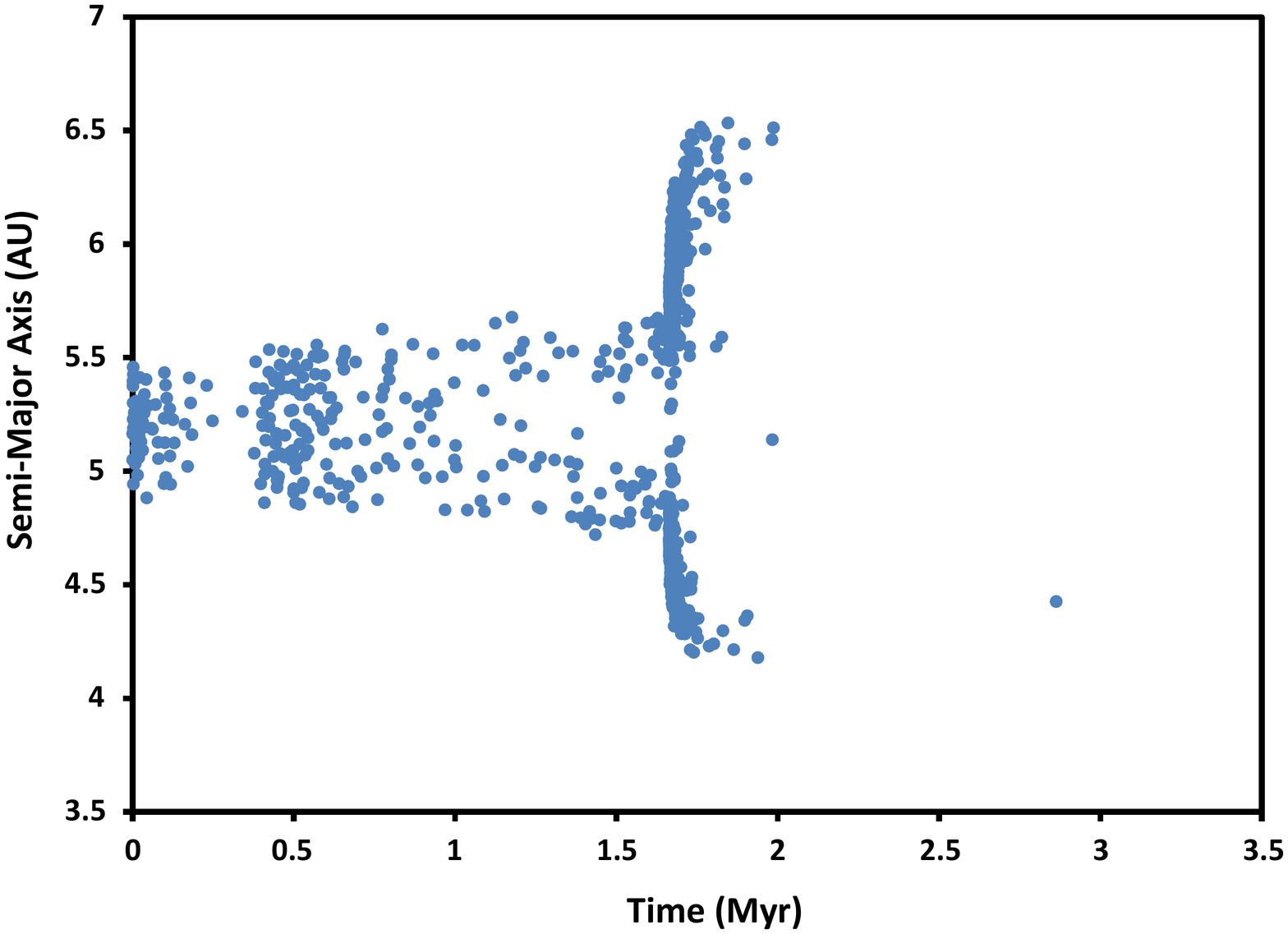}
\hskip -40pt
\includegraphics[scale=0.38]{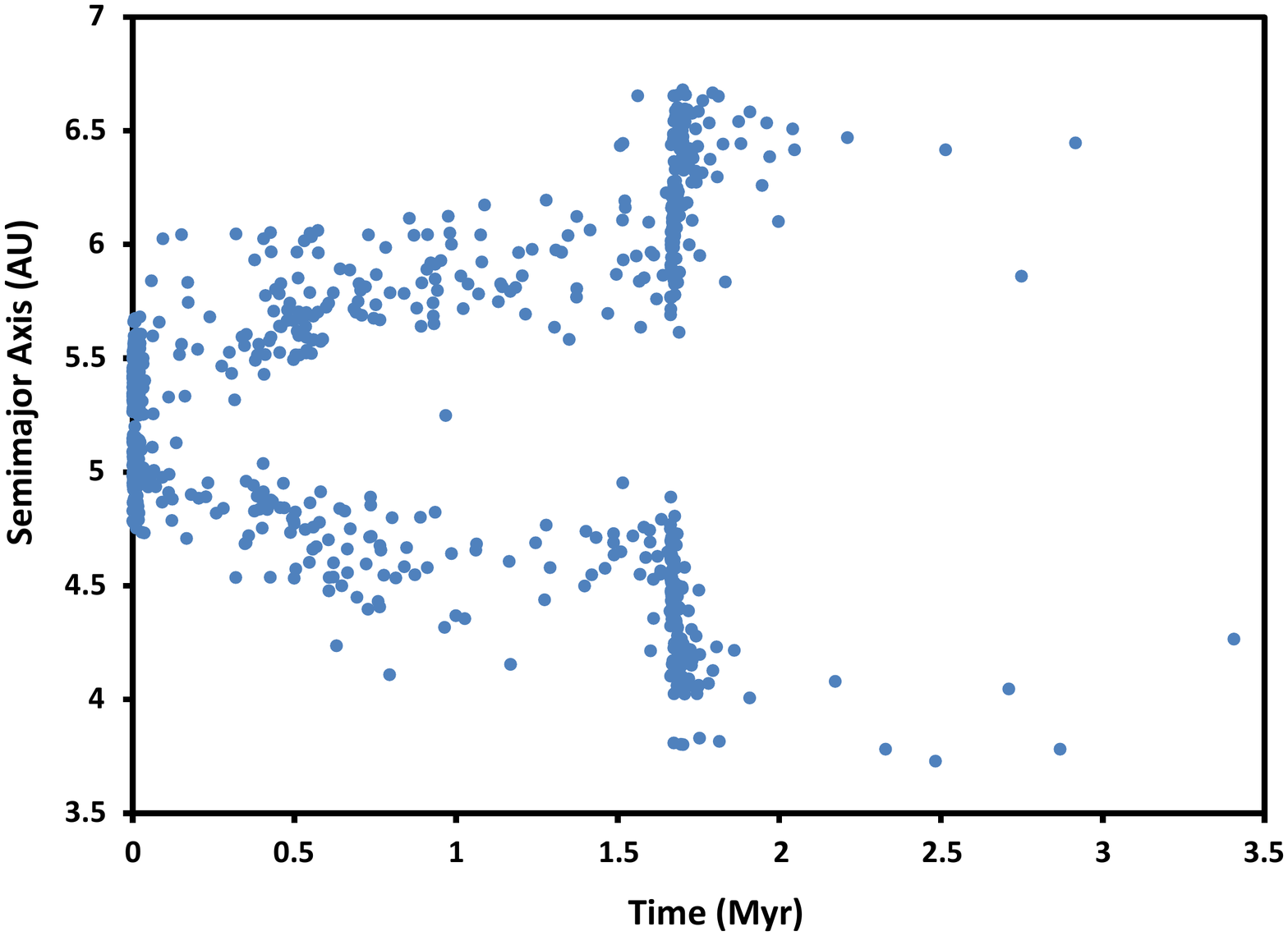}
\vskip -40pt
\hskip -21pt
\includegraphics[scale=0.38]{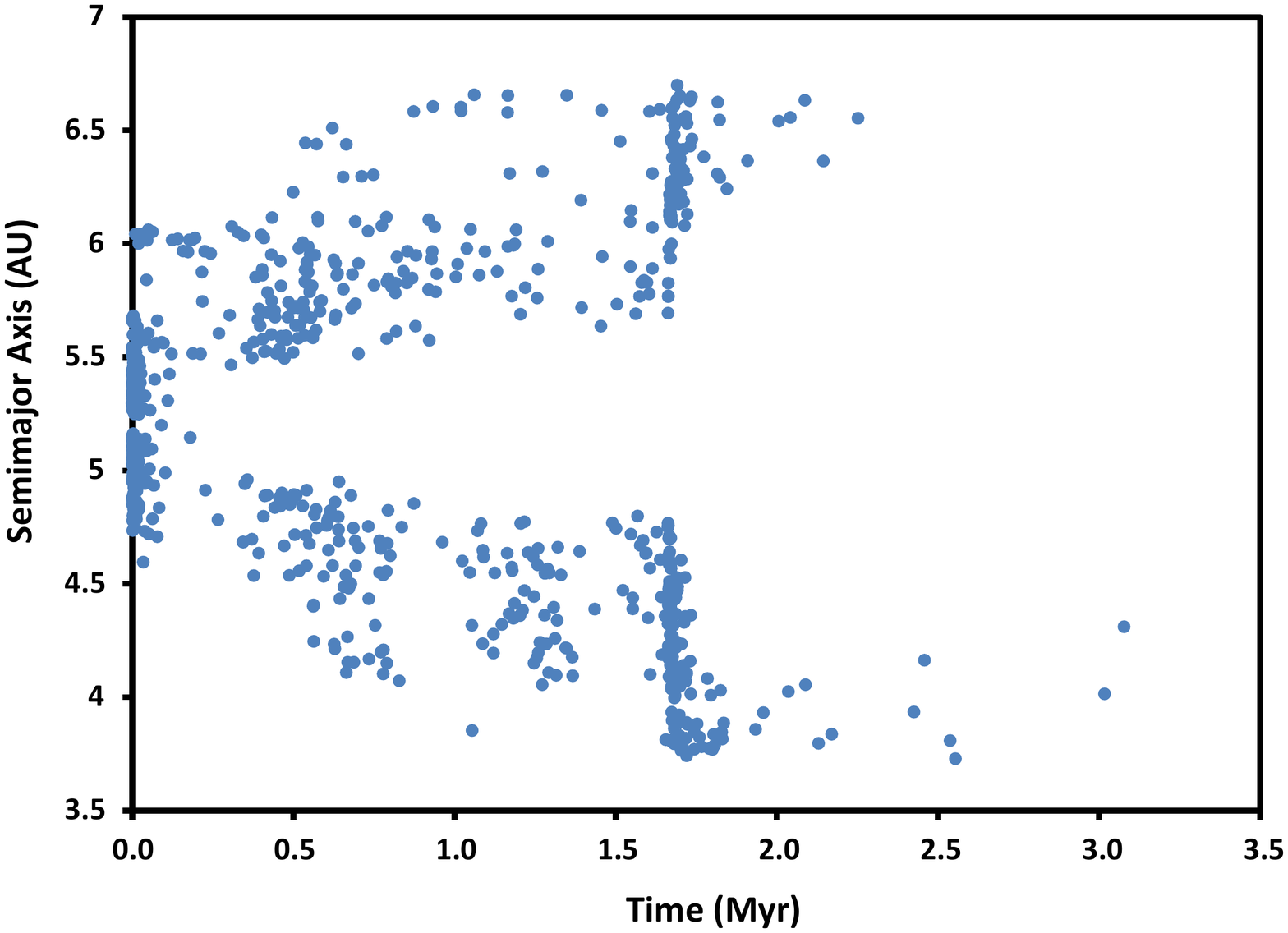}
\hskip -40pt
\includegraphics[scale=0.38]{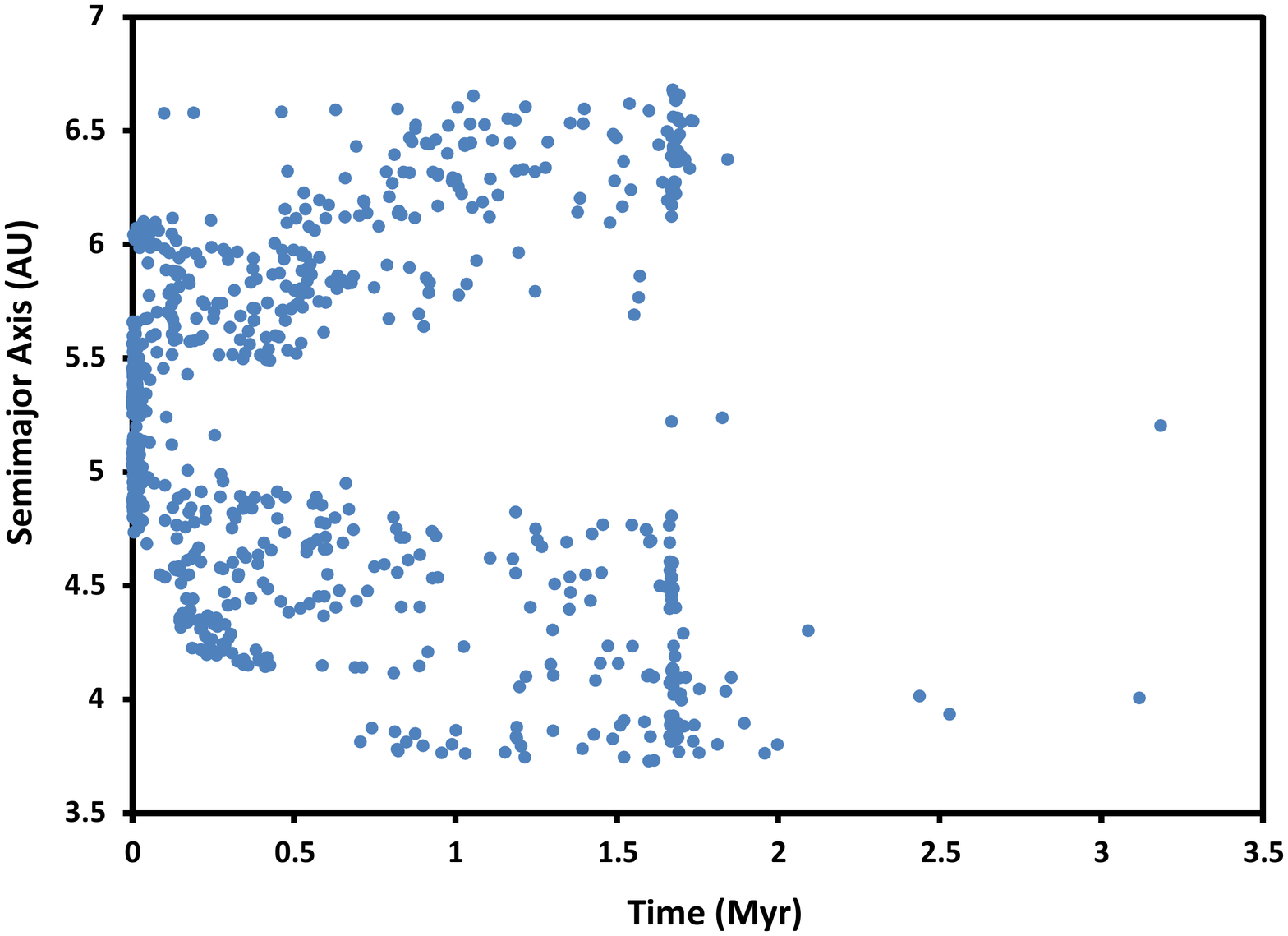}
\vskip -20pt
\caption{The tuning-fork graphs of the accretion times of 100 km rocky planetesimals. The top-right panel corresponds 
to 1/10 of Saturn's mass. Bottom-left panel is for the system with 1/3 of the mass of Saturn, and the bottom-right panel
shows the results for the system with full Saturn-mass.}
\label{fig10}
\end{figure*}

\section{The effect of Saturn}
An analysis of the orbital dynamics of the planetesimals in our integrations indicates that, compared to the systems
of Paper-I, more planetesimals in our systems were left un-accreted and/or were ejected from the system. Figure 9 shows
this for the case of 100 km pure-rock planetesimals. Shown here are the final eccentricities of the surviving 
planetesimals (in red) compared to their initial eccentricities (in black) as a function of their initial semimajor axes. 
The top-left panel corresponds 
to the integrations in Paper-I and the three other panels show the results for our integrations with the mass of the
planet in the orbit of Saturn equal to 1/10 (top-right), 1/3 (bottom-left), and full Saturn-mass (bottom -right). 
As shown here, in integrations without Saturn, all planetesimals in the region between 5 AU and 5.5 AU were accreted 
by the growing proto-Jupiter. While there is a small number of scattered planetesimals, many remained in the system 
maintaining their original eccentricities (i.e., remain unaffected by the growing Jupiter). However, Once a planet
is introduced in the orbit of Saturn, the perturbation of this planet causes the orbits of more planetesimals to
become highly eccentric and less number of planetesimals in the vicinity of Jupiter to be accreted. As expected,
the scattering becomes stronger for larger values of Saturn's mass.

In addition to increasing their orbital eccentricities, interaction with Saturn also increases the velocities with
which planetesimals encounter Jupiter's envelope. This increase in the encounter velocity, enhances the 
effects of gas-drag and ram-pressure, thereby, reducing the time of accretion. Figure 10 shows the {\it tuning fork} diagrams
corresponding to the accretion time of fully captured planetesimals in term of their initial semimajor axis. Once again, 
the top-left panel shows the results for the case without Saturn, and the three subsequent panels represent results for
the cases of 1/10 (top-right), 1/3 (bottom-left), and full Saturn-mass (bottom-right). As shown here, when the mass 
of Saturn increases, more planetesimals are captured at earlier times. However, irrespective of the mass of Saturn, 
accretion peaks at approximately 1.7 Myr for those planetesimals in the close vicinity of Jupiter.
The latter is not unexpected as it corresponds to the time of the collapse of the envelope where the increase in the envelope 
gas density enhances the rate of accretion.

\section{Summary and concluding remarks}
We have studied the interactions of planetesimals with the gaseous envelope of the growing Jupiter in the 
core-accretion model. Our study aims at calculating a more accurate mass accretion rate for the growing
giant planet and advances our previous calculations (Paper-I) by including the gravitational perturbation 
of Saturn. 

Using our special purpose integrator ESSTI, we tracked the orbits of planetesimals inside the 
envelope and calculated their mass-loss due to the heating by gas-drag and fragmentation caused by 
ram-pressure. Results confirmed our previous findings that, unlike the reports by \citet{Inaba03},
in general, ram-pressure is the main process that affects the orbits of planetesimals when they first encounter 
the envelope. The high encounter velocities of these objects, especially when Saturn is included, triggers 
ram-pressure, causing the body to break up into small pieces and subsequently be accreted as its small
fragments vaporize due ablation while descending in the envelope.

An analysis of the efficiency of the breakage and vaporization of planetesimals with different material 
compositions indicated that mix-composition planetesimals have the largest contribution to the metalicity 
of the envelope and even the growth of the planet's core. As these planetesimals disintegrate, their ice components 
vaporize almost immediately. Their rock components, however, dissolve gradually as they descend toward the 
center of the planet, enhancing the envelope's metalicity. Integrations show that some of these
rocky fragments survive the descent and collide with the surface of the core contributing to its growth as well. 

Our analysis also shows that the rate of collision of fragments with the core increases with Saturn's mass (Table 3). 
This is an expected result that has to do with the fact that a larger planet perturbs the orbits of planetesimals more 
strongly increasing their encounter velocities and enhancing the fragmentation effect of the ram-pressure. The latter 
results in producing more fragments with larger post-break up velocities. 

Among the three types of compositions considered here, pure-rock planetesimals show a more interesting and complicated 
motion. The high density of these objects prevents them from 
being disintegrated when they first encounter the envelope. As a result, these planetesimals pass through the envelope 
multiple times losing mass to vaporization until their internal structure becomes weak enough for ram-pressure 
to break them apart. Results of integrations for 1-100 km-sized rocky bodies show that the full accretion occurs
after approximately  $5\times 10^5$\, years and almost entirely deep in the envelope where the temperature 
is higher than 2700-3000 K. 

A comparison between the results presented here and those of Paper-I demonstrates that when Saturn is included,
the rate of mass-accretion becomes smaller. The gravitational perturbation of Saturn increases the orbital
eccentricities of many of the planetesimals, scattering them to the regions outside the influence zone 
of Jupiter, thereby reducing the number of bodies that encounter Jupiter's envelope.
For those planetesimals that enter the envelope, the increase in orbital eccentricity appears as an
increase in their encounter velocities. The latter enhances the efficiency of gas-drag and ram-pressure
in dissolving the body, and shortens the time of 
its accretion. Because the perturbation of Saturn is a direct function of its mass, these findings
imply that the rate of the accretion of planetesimals by Jupiter's envelope was higher when Saturn was 
small and decreased with time as Saturn became larger (and its gravitational perturbation became stronger).

That the rate of accretion decreases with Saturn's mass has a direct consequence on the final size of Jupiter's
core. Results of Paper-I suggested that during the first $5\times 10^5$\, years, the rate of mass accretion  
was lower than that assumed by \cite{Lozovsky17}. When the perturbation of Saturn is included, that rate becomes 
even smaller. A smaller core, in turn, implies a smaller and less massive envelope meaning that the envelope
considered here might not have been of a fully realistic size. A self-consistent model requires that the
calculations, in addition to the effect of Saturn on the motion of planetesimals, to include the effect 
of Saturn's growth on Jupiter's envelope, as well.

As mentioned in Section 2, to maintain focus on Saturn's effects, we did not consider the interaction 
of planetesimals with the nebular gas. This interaction, which appears in different forms, can affect 
the motion of planetesimals prior to their encounter with the envelope. For instance, as shown by 
\citet{Zhou07}, \citet{Shiraishi08} and \citet{Shibata19}, the drag force of the nebula will change the encounter 
velocities of planetesimals which may, at least partially, counter the perturbing effect of 
Saturn by preventing the orbital eccentricities of planetesimals from reaching very high values. The latter 
may slightly improve the accretion rate of these objects by reducing the number of scattered planetesimals and 
increasing the rate of their encounter with the envelope. 
Including this effect is the subject of future studies.

Finally, it is important to note that to ensure that our simulations would start from
similar initial conditions as those in paper I, so that a comparison between the results of the two
studies would be meaningful, we only included planetesimals with initial 
semimajor axes between 3.7 (AU) and 6.7 (AU). In paper I, this range was sufficient because 
only Jupiter was included. Saturn, however, will affect planetesimals further out, and some 
of these planetesimals may enter Jupiter's envelope. In future calculations, the  
initial distribution of planetesimals will be extended to farther distances.

\section*{Acknowledgments}
Support from NASA grants 80NSSC18K0519 and 80NSSC21K1050, and NSF grant AST-2109285 for NH is acknowledged.


\begin{thebibliography}{}

\bibitem[Chambers (1999)]{Chambers99}
Chambers, J. E. 1999, MNRAS, 304, 793

\bibitem[Alibert et al (2005)]{Alibert05}
Alibert, Y., Mordasini, C., Benz, W. \& Winisdoerffer, C. 2005, A\&A, 434, 343

\bibitem[D' Angleo et al. (2014)]{Dangelo14}
D’Angelo, G., Weidenschilling, S. J., Lissauer, J. J. \& Bodenheimer, P. 2014, Icar, 241, 298

\bibitem[Greenzweig \& Lissauer (1990)]{Greenzweig90}
Greenzweig, Y. \& Lissauer, J. J. 1990, Icar, 87, 40

\bibitem[Greenzweig \& Lissauer (1992)]{Greenzweig92}
Greenzweig, Y. \& Lissauer, J. J. 1992, Icar, 100, 440

\bibitem[Haghighipour \& Scott (2012)]{Haghighipour12}
Haghighipour, N. \& Scott, E. R. D. 2012, ApJ, 749, 113

\bibitem[Haghighipour \& Winter (2016)]{Haghighipour16}
Haghighipour, N. \& Winter, O. C. 2016, Celest. Mech. Dyn. Astr., 124, 235

\bibitem[Iaroslavitz \& Podolak (2007)]{Iaros07}
Iaroslavitz, E. \& Podolak, M. 2007, Icar, 187, 600

\bibitem[Ida \& Makino (1993)]{Ida93}
Ida, S. \& Makino, J. 1993, Icarus, 106, 210

\bibitem[Inaba et al. (2001)]{Inaba01}
Inaba, S., Tanaka, H., Nakazawa, K., Wetherill, G. W. \& Kokubo, E. 2001, Icarus, 149, 235

\bibitem[Inaba \& Ikoma (2003)]{Inaba03}
Inaba, S. \& Ikoma, M. 2003, A\&A, 410, 711

\bibitem[Lozovsky et al. (2017)]{Lozovsky17}
Lozovsky, M., Helled, R., Rosenberg, E. D., \& Bodenheimer, P. 2017, ApJ, 836, 227

\bibitem[Movshovitz et al. (2010)]{Movsh10}
Movshovitz, N., Bodenheimer, P., Podolak, M. \& Lissauer, J. J. 2010, Icar, 209, 616

\bibitem[Nakazawa et al. (1989)]{Nakazawa89}
Nakazawa, K., Ida, S. \& Nakagawa, Y. 1989, A\&A, 220, 293

\bibitem[Podolak et al. (1988)]{Podolak88}
Podolak, M., Pollack, J. B., \& Reynolds, R. T. 1988, Icar, 73, 163

\bibitem[Podolak et al. (2020)]{Podolak20}
Podolak, M., Haghighipour, N., Bodenheimer, P., Helled, R. \& Podolak, E. 2020, ApJ, 899, 45

\bibitem[Pollack et al. (1979)]{Pollack79}
Pollack, J. B., Burns, J. A., \& Tauber, M. E. 1979, Icar, 37, 587

\bibitem[Pollack et al. (1996)]{Pollack96}
Pollack, J. B., Hubickyj, O., Bodenheimer, P., Lissauer, J. J., Podolak, M. \& Greenzweig, Y.
1996, Icar, 124, 62

\bibitem[Shibata \& Ikoma (2019)]{Shibata19}
Shibata, S. \& Ikoma, M. 2019, MNRAS, 487, 4510

\bibitem[Shibata et al. (2020)]{Shibata20}
Shibata, S., Helled, R. \& Ikoma, M. 2020, A\&A, 633, id.A33

\bibitem[Shibata et al. (2022)]{Shibata22}
Shibata, S., Helled, R. \& Ikoma, M. 2020, A\&A, 659, id.A28

\bibitem[Shiraishi \& Ida (2008)]{Shiraishi08}
Shiraishi, M. \& Ida, S. 2008, ApJ, 684, 1416

\bibitem[Venturini et al. (2016)]{Venturini16}
Venturini, J., Alibert, Y. \& Benz, W. 2016, A\&A, 596, A90

\bibitem[Zhou \& Lin (2007)]{Zhou07}
Zhou, J-l \& Lin, D. N. C. 2007, ApJ, 666, 447

\end{thebibliography}
\end{document}